\documentclass[amsmath,aps,showpacs,a4paper,10pt]{revtex4}

 \usepackage{epsf}
 \usepackage{graphicx}    

\begin{document}

 \newcommand{\be}[1]{\begin{equation}\label{#1}}
 \newcommand{\ee}{\end{equation}}
 \newcommand{\bea}{\begin{eqnarray}}
 \newcommand{\eea}{\end{eqnarray}}
 \def\disp{\displaystyle}


 \newcommand{\dmunit}{ {\rm pc\hspace{0.24em} cm^{-3}} }
 \newcommand{\cmn}{,\,}

 \begin{titlepage}

 \begin{flushright}
 arXiv:2301.08194
 \end{flushright}

 \title{\Large \bf Fast Radio Bursts as Standard Candles for Cosmology}

 \author{Han-Yue~Guo\,}
 \email[\,email address:\ ]{guohanyue7@163.com}
 \affiliation{School of Physics,
 Beijing Institute of Technology, Beijing 100081, China}

 \author{Hao~Wei\,}
 \email[\,Corresponding author;\ email address:\ ]{haowei@bit.edu.cn}
 \affiliation{School of Physics,
 Beijing Institute of Technology, Beijing 100081, China}

 \begin{abstract}\vspace{1cm}
 \centerline{\bf ABSTRACT}\vspace{2mm}
 Recently, fast radio bursts (FRBs) have become a thriving field in
 astronomy and cosmology. Due to their extragalactic and cosmological
 origin, they are useful to study the cosmic expansion and the intergalactic
 medium (IGM). In the literature, the dispersion measure (DM) of FRB has
 been considered extensively. It could be used as an indirect proxy of
 the luminosity distance $d_L$ of FRB. The observed DM contains the
 contributions from the Milky Way (MW), the MW halo, IGM, and the host
 galaxy.~Unfortunately, IGM and the host galaxy of FRB are poorly known
 to date, and hence the large uncertainties of $\rm DM_{IGM}$
 and $\rm DM_{host}$ in DM plague the FRB cosmology.~Could we avoid
 DM in studying cosmology? Could we instead consider the luminosity
 distance $d_L$ directly in the FRB cosmology? We are interested to find
 a way out for this problem in the present work. From the lessons of
 calibrating type~Ia supernovae (SNIa) or long gamma-ray bursts (GRBs)
 as standard candles, we consider a universal subclassification scheme
 for FRBs, and there are some empirical relations for them. In
 the present work, we propose to calibrate type~Ib FRBs as standard
 candles by using a tight empirical relation without DM. The calibrated
 type~Ib FRBs at high redshifts can be used like SNIa to constrain the
 cosmological models.~We also test the key factors affecting the calibration
 and the cosmological constraints. Notice that the results are speculative,
 since they are based on simulations for the future (instead of now). We
 stress that the point in the present work is just to propose a viable
 method and a workable pipeline using type~Ib FRBs as standard candles
 for cosmology in the future.
 \end{abstract}


 \maketitle

 \end{titlepage}

 \renewcommand{\baselinestretch}{1.0}


\section{Introduction}\label{sec1}

Recently, fast radio bursts (FRBs) have become a thriving field in astronomy
 and cosmology~\cite{NAFRBs,Lorimer:2018rwi,Keane:2018jqo,Petroff:2021wug,
 Zhang:2022uzl,Zhang:2020qgp,Xiao:2021omr}. FRBs are millisecond-duration
 transient radio sources, and their origin is still unknown to date. Most of
 them are at extragalactic/cosmological distances, as suggested by their
 large dispersion measures~(DMs) well in excess of the Galactic values.
 Therefore, it is of interest to study cosmology and the intergalactic
 medium (IGM) with FRBs~\cite{NAFRBs,Lorimer:2018rwi,Keane:2018jqo,
 Petroff:2021wug,Zhang:2022uzl,Zhang:2020qgp,Xiao:2021omr}.

To date, there are many FRB theories in the literature~\cite{NAFRBs,
 Lorimer:2018rwi,Keane:2018jqo,Petroff:2021wug,Zhang:2022uzl,Zhang:2020qgp,
 Xiao:2021omr,Platts:2018hiy}. For example, FRBs might originate from
 magnetars, mergers/interactions of compact objects (such as neutron
 star (NS), white dwarf (WD), black hole (BH), strange star (SS), axion
 star), collapse of compact objects, active galactic nuclei
 (AGN), supernovae remnants (SNR), starquakes, giant pulses, superradiance,
 collision/interaction between NS and asteroids/comets, cosmic
 strings, cosmic combs, pulsar lightnings, variable stars, wandering
 pulsar beams. We strongly refer to~\cite{Platts:2018hiy} for the living
 FRB theory catalogue.

As is well known, one of the key observational quantities of FRBs is the
 dispersion measure (DM), namely the column density of the free
 electrons, due to the ionized medium (plasma) along the path. Since the
 distance along the path in DM records the expansion history of the
 universe, it plays a key role in the FRB cosmology. The observed DM of
 FRB at redshift $z$ can be separated into~\cite{Deng:2013aga,
 Yang:2016zbm,Gao:2014iva,Zhou:2014yta,Qiang:2019zrs,Qiang:2020vta,
 Qiang:2021bwb,Qiang:2021ljr,Guo:2022wpf}
 \be{eq1}
 {\rm DM_{obs}=DM_{MW}+DM_{halo}+DM_{IGM}+DM_{host}}/(1+z)\,,
 \ee
 where $\rm DM_{MW}$, $\rm DM_{halo}$, $\rm DM_{IGM}$, $\rm DM_{host}$ are
 the contributions from the Milky Way (MW), the MW halo, IGM, the host
 galaxy (including interstellar medium of the host galaxy and
 the near-source plasma), respectively. Obviously, the main contribution to
 DM of FRB comes from IGM. The mean of $\rm DM_{IGM}$ is given
 by~\cite{Deng:2013aga,Yang:2016zbm,Gao:2014iva,Zhou:2014yta,
 Qiang:2019zrs,Qiang:2020vta,Qiang:2021bwb,Qiang:2021ljr,Guo:2022wpf}
 \be{eq2}
 \langle{\rm DM_{IGM}}\rangle=\frac{3cH_0\Omega_b}{8\pi G m_p}
 \int_0^z\frac{f_{\rm IGM}(\tilde{z})\,f_e(\tilde{z})\left(1+
 \tilde{z}\right)d\tilde{z}}{h(\tilde{z})}\,,
 \ee
 where $c$ is the speed of light, $H_0$ is the Hubble constant, $\Omega_b$
 is the present fractional density of baryons, $G$ is the gravitational
 constant, $m_p$ is the mass of proton, $h(z)\equiv H(z)/H_0$ is the
 dimensionless Hubble parameter, $f_{\rm IGM}(z)$ is the fraction of baryon
 mass in IGM, and $f_e(z)$ is the ionized electron number fraction per
 baryon. The latter two are functions of redshift $z$ in principle.

One can see that $\langle{\rm DM_{IGM}}\rangle$ is related to the expansion
 history of the universe through $h(z)$ according to Eq.~(\ref{eq2}). Note
 that $\langle{\rm DM_{IGM}}\rangle$ is the mean value of $\rm DM_{IGM}$ in
 all directions of the lines of sight. This is only valid due to the
 well-known cosmological principle, which assumes that the universe is
 homogeneous and isotropic on cosmic/large scales. However, it is not
 homogeneous and isotropic on local/small scales. Thus, ${\rm DM_{IGM}}
 \not=\langle{\rm DM_{IGM}}\rangle$ for most observed FRB, since the
 plasma density fluctuates along the line of sight~\cite{McQuinn:2013tmc,
 Ioka:2003fr,Inoue:2003ga,Jaroszynski:2018vgh} (see also
 e.g.~\cite{Deng:2013aga,Yang:2016zbm,Gao:2014iva,Zhou:2014yta,
 Qiang:2019zrs,Qiang:2020vta,Qiang:2021bwb,Qiang:2021ljr,Guo:2022wpf}). In
 the literature, the deviation of $\rm DM_{IGM}$ from $\langle{\rm DM_{IGM}}
 \rangle$ is usually characterized by the uncertainty $\sigma_{\rm IGM}(z)$,
 which is a function of redshift $z$ in principle. Unfortunately, IGM is
 poorly known in fact. So, one can only consider various empirical
 $\sigma_{\rm IGM}(z)$ in the literature, which is usually large ($\sim
 {\cal O}(10^2)$ to ${\cal O}(10^3)\;\dmunit$)~\cite{McQuinn:2013tmc,
 Ioka:2003fr,Inoue:2003ga,Jaroszynski:2018vgh} (see also
 e.g.~\cite{Deng:2013aga,Yang:2016zbm,Gao:2014iva,Zhou:2014yta,
 Qiang:2019zrs,Qiang:2020vta,Qiang:2021bwb,Qiang:2021ljr,Guo:2022wpf}).
 In fact, the large uncertainty $\sigma_{\rm IGM}(z)$ in $\rm DM_{IGM}$
 is one of the main troubles in the FRB cosmology. On the other hand,
 $f_{\rm IGM}(z)$ and $f_e(z)$ in $\langle{\rm DM_{IGM}}\rangle$ are both
 functions of redshift $z$ in principle. Since we are rather ignorant of
 IGM and the cosmic ionization history, $f_{\rm IGM}$ and $f_e$ have
 been extensively assumed to be constant in the literature. Thus, this also
 leads to uncertainties in the FRB cosmology.

For convenience, one can introduce the extragalactic DM~\cite{Deng:2013aga,
 Yang:2016zbm,Gao:2014iva,Zhou:2014yta,Qiang:2019zrs,Qiang:2020vta,
 Qiang:2021bwb,Qiang:2021ljr,Guo:2022wpf},
 namely
 \be{eq3}
 {\rm DM_E=DM_{obs}-DM_{MW}-DM_{halo}=DM_{IGM}+DM_{host}}/(1+z)\,.
 \ee
 In the literature, $\rm DM_E$ has been extensively used to study cosmology.
 Note that the host galaxy of FRB and hence its contribution to DM are also
 poorly known. So, $\rm DM_{host}$ was usually assumed to be constant or
 was randomly assigned from some simplified distributions in the literature
 \cite{Deng:2013aga,Yang:2016zbm,Gao:2014iva,Zhou:2014yta,Qiang:2019zrs,
 Qiang:2020vta,Qiang:2021bwb,Qiang:2021ljr,Guo:2022wpf}.~Of course,
 these simplifications prevent us from better understanding
 the cosmic expansion history. Here we refer to e.g. the highly cited
 Ref.~\cite{Jaroszynski:2018vgh} for a qualitative impression on the
 cosmological constraints. In~\cite{Jaroszynski:2018vgh}, using
 $\rm DM_E$ of 100 simulated FRBs to constrain the flat $\Lambda$CDM model,
 $\Omega_m=0.29\pm 0.10$ and $0.26\pm 0.05$ were found for the cases of
 $z_{\rm max}=1.5$ and $2.5$, respectively. To improve the cosmological
 constraints, it is common to combine FRBs with other observations such
 as type Ia supernovae (SNIa), cosmic microwave background (CMB), and
 baryon acoustic oscillations (BAO) in the literature. However, such
 improvements are mainly dominated by SNIa, CMB and BAO, rather than
 FRBs themselves.

Although DM is extensively used to study cosmology in the literature, it is
 just an indirect proxy of the luminosity distance $d_L$ of FRB. In
 addition, the large uncertainties of $\rm DM_{IGM}$ and $\rm DM_{host}$ in
 DM plague the FRB cosmology, as mentioned above. Could we avoid DM in
 studying cosmology? Could we instead consider the luminosity distance $d_L$
 directly in the FRB cosmology? We are interested to find a way out for this
 problem in the present work.

Of course, it is difficult to obtain the luminosity distance $d_L$ of FRB.
 If this FRB is well localized to a host galaxy, and the luminosity distance
 of this galaxy has been well measured, $d_L$ of this FRB is equal to the
 one of its host galaxy. Alternatively, if the luminosity distance of
 objects in the same host galaxy of an FRB (e.g.~supernova, gamma-ray burst,
 or gravitational wave event, not necessarily counterparts) can be measured,
 $d_L$ of this FRB is also on hand. But these make sense only at low
 redshifts. We should find a new method to obtain the luminosity distances
 $d_L$ for FRBs at high redshifts.

History always repeats itself. As is well known, the distance ladder has
 been used in many neighboring fields. For example, the luminosity distances
 of Cepheids can be obtained by using the empirical period versus luminosity
 ($P-L$, or, Leavitt) relation~\cite{Leavitt:1912}. If a type~Ia supernova
 (SNIa) shares the same host galaxy with a Cepheid, the luminosity distance
 of this SNIa is equal to the one of Cepheid. The empirical SNIa light-curve
 versus luminosity (Phillips) relation~\cite{Phillips:1993ng} could be
 calibrated by using these SNIa at low redshifts.~Thus, the luminosity
 distances of SNIa at high redshifts can be obtained by extending the
 calibrated SNIa Phillips relation~\cite{Riess:2021jrx}. Similar to the case
 of calibrating SNIa as secondary standard candles by using Cepheids as
 primary standard candles, this has also been considered in the field of
 gamma-ray bursts (GRBs). The luminosity distances of long GRBs at low
 redshifts can be obtained by using e.g. interpolation from the ones of
 SNIa at nearby redshifts~\cite{Liang:2008kx,Wei:2010wu}. Various empirical
 relations, especially the $E_{\rm p\cmn i}-E_{\rm iso}$ (Amati) relation
 \cite{Amati:2002ny}, could be calibrated by using these long GRBs at low
 redshifts~\cite{Schaefer:2006pa} (see also e.g.~\cite{Liang:2008kx,
 Wei:2010wu}). Thus, the luminosity distances of long GRBs at high redshifts
 can be obtained by extending the calibrated GRB empirical relation(s)
 \cite{Liang:2008kx,Wei:2010wu,Wei:2008kq,Liu:2014vda}.

The lessons from the above history of calibrating SNIa or long GRBs
 as standard candles are (a) at least a tight empirical relation involving
 the luminosity distance is necessary; (b) the empirical relations are only
 valid for some subclasses, namely SNIa (not for type Ib, Ic and II
 supernovae), or long GRBs (not for short GRBs). In the present work, we
 try to calibrate FRBs as standard candles. Thus, we should introduce
 a universal subclassification scheme for FRBs, and find a tight empirical
 relation involving the luminosity distance for a subclass of
 FRBs. Fortunately, they are ready in our previous works.

This paper is organized as followings. In Sec.~\ref{sec2}, we briefly
 introduce the universal subclassification scheme and the empirical
 relations for FRBs. In Sec.~\ref{sec3}, we describe the method to calibrate
 type~Ib FRBs as standard candles. In Sec.~\ref{sec4}, we test the key
 factors affecting the calibration and the cosmological constraints. In
 Sec.~\ref{sec5}, some brief concluding remarks are given.


\section{Subclassification and empirical relations for FRBs}\label{sec2}

One of the interesting topics is the classification of FRBs, which is
 closely related to the origin of FRBs. Clearly, different physical
 mechanisms are required by different classes of FRBs. Well motivated by
 the actual observations, they are usually classified into two populations:
 non-repeating FRBs and repeating FRBs~\cite{NAFRBs,Lorimer:2018rwi,
 Keane:2018jqo,Petroff:2021wug,Zhang:2022uzl,Zhang:2020qgp,Xiao:2021omr}.
 For convenience, we call them type I and II
 FRBs in~\cite{Guo:2022wpf}, respectively.

In the literature, it was speculated for a long time that
 the FRB distribution tracks the cosmic star formation
 history~(SFH)~\cite{NAFRBs,Lorimer:2018rwi,Keane:2018jqo,Petroff:2021wug,
 Zhang:2022uzl,Zhang:2020qgp,Xiao:2021omr}. The well-known Galactic FRB
 200428 associated with the young magnetar SGR 1935+2154
 \cite{Andersen:2020hvz,Bochenek:2020zxn,Lin:2020mpw,Li:2020qak} confirmed
 that some FRBs originate from young magnetars. On the other hand, some
 repeating FRBs (such as FRB 121102, FRB 180916.J0158+65, FRB 20190520B,
 FRB 20181030A) were observed to be closely correlated
 with star-forming activities~\cite{Tendulkar:2017vuq,Marcote:2020ljw,
 Niu:2021bnl,Bhardwaj:2021hgc}. Therefore, it is reasonable to expect
 that at least some (if not all) FRBs are associated with young stellar
 populations, and hence their distribution tracks SFH. However, this
 speculation was challenged recently. The well-known repeating
 FRB 20200120E in a globular cluster of the nearby galaxy M81
 \cite{Bhardwaj:2021xaa,Kirsten:2021llv,Nimmo:2021yob} suggested that some
 FRBs are associated with old stellar populations instead. In addition, the
 host of non-repeating FRB 20210117A was also found to be a dwarf galaxy
 with little on-going star formation~\cite{Bhandari:2022ton}. The host
 of FRB 20220509G is a red, old, massive elliptical galaxy with low star
 formation rate~\cite{DeepSynopticArrayTeam:2023fxs}. We refer
 to e.g.~\cite{Michilli:2022bbs,Law:2023ibd} for more examples requiring
 FRB progenitor formation channels associated with old stellar populations.
 On the other hand, it was claimed in~\cite{Zhang:2021kdu} that the 536
 bursts of the first CHIME/FRB catalog~\cite{CHIMEFRB:2021srp} (released
 in June 2021, including 474 one-off bursts and 62 repeat bursts from 18
 repeaters observed at the central frequency $\nu_c=600\,{\rm MHz}$)
 as a whole do not track SFH. In~\cite{Qiang:2021ljr}, it was
 independently confirmed that the FRB distribution tracking SFH
 can be rejected at high confidence, and a suppressed evolution (delay)
 with respect to SFH was found.


\begin{table}[tb]
 \renewcommand{\arraystretch}{1.0}
 \begin{center}
 \vspace{-5.5mm}  
 \hspace{-5mm}  
 \begin{tabular}{c|c|c} \hline\hline
 {\bf FRBs} & \begin{tabular}{c} $\vspace{-3.7mm}$\\ {\bf Class (a)\;:} \\ \ associated with old stellar populations \\[1.41mm] \end{tabular}
 & \begin{tabular}{c} {\bf Class (b)\;:} \\ \ associated with young stellar populations \end{tabular} \\ \hline
 \begin{tabular}{c} {\bf Type I\;:} \\ Non-repeating\ \ \\ \end{tabular}
 & \begin{tabular}{c} $\vspace{-3.7mm}$\\ {\bf Type Ia\;:} \\ Non-repeating FRBs \\ associated with old stellar populations \\[1.41mm] \end{tabular}
 & \begin{tabular}{c} {\bf Type Ib\;:} \\ Non-repeating FRBs \\ \ associated with young stellar populations\ \end{tabular} \\ \hline
 \begin{tabular}{c} {\bf Type II\;:} \\ Repeating\ \ \end{tabular}
 & \begin{tabular}{c} $\vspace{-3.7mm}$\\ {\bf Type IIa\;:} \\ Repeating FRBs \\ associated with old stellar populations \\[1.41mm] \end{tabular}
 & \begin{tabular}{c} {\bf Type IIb\;:} \\ Repeating FRBs \\ \ associated with young stellar populations \end{tabular} \\ \hline \hline
 \end{tabular}
 \end{center}
 \vspace{-0.5mm}  
 \caption{\label{tab1} A universal subclassification scheme for FRBs
 proposed in~\cite{Guo:2022wpf}.}
 \end{table}


The above discussions based on the actual observations have motivated us
 in~\cite{Guo:2022wpf} to speculate that some FRBs are associated with young
 populations and hence their distribution tracks SFH, while the other FRBs
 are associated with old populations and hence their distribution does not
 track SFH. This led us to propose a universal subclassification scheme
 for FRBs~\cite{Guo:2022wpf}, as shown in Table~\ref{tab1}.

In~\cite{Guo:2022wpf}, we have conducted this subclassification scheme
 for FRBs by using the actual data of the first CHIME/FRB
 catalog~\cite{CHIMEFRB:2021srp}. These FRBs are subclassified in the
 transient duration $\nu W$ versus spectral luminosity $L_\nu$ phase
 plane, with some isothermal lines of brightness temperature $T_B$.
 The $\nu W-L_\nu$ phase plane has been divided into ten regions by
 three dividing lines $\nu W=10^{-3}\,{\rm GHz\;s}$, $L_\nu=10^{34}\,
 {\rm erg/s/Hz}$ and $T_B=2\times 10^{35}\,{\rm K}$, which are determined
 by the minimal p-values of Kolmogorov-Smirnov (KS) test~\cite{Guo:2022wpf}.
 Note that there are 430 non-repeating (type~I) FRBs after the robust cut
 in the first CHIME/FRB catalog~\cite{CHIMEFRB:2021srp}. Then, we test
 these 10 regions one by one. For each region, 430 type~I FRBs are divided
 into two samples inside or outside this region. We compare their redshift
 distributions by using KS test, and also check whether one of these two
 sample tracks SFH by using the method proposed in~\cite{Qiang:2021ljr,
 Zhang:2021kdu,Zhang:2020ass}. Finally, we find that region~(8) is very
 successful, and hence the physical criteria for the subclassification
 of type~I FRBs have been clearly determined~\cite{Guo:2022wpf}, namely
 \bea
 &{\rm Type~Ia:}\quad & L_\nu\leq 10^{34}\,{\rm erg/s/Hz}\quad
 \&\quad T_B\geq 2\times 10^{35}\,{\rm K}\,,\label{eq4}\\[1mm]
 &{\rm Type~Ib:}\quad & {\rm otherwise}\,.\label{eq5}
 \eea
 Note that these physical criteria are suitable for the first CHIME/FRB
 catalog~\cite{CHIMEFRB:2021srp}, and they might be changed for the larger
 and better FRB datasets in the future, but the universal subclassification
 scheme given in Table~\ref{tab1} will always hold. We find that in the
 first CHIME/FRB catalog~\cite{CHIMEFRB:2021srp}, 65 type~Ia FRBs do not
 track SFH, but 365 type~Ib FRBs do track SFH at high confidence. Similarly,
 we speculate that the possible physical criteria for the
 subclassification of type~II FRBs might be given by~\cite{Guo:2022wpf}
 \bea
 &{\rm Type~IIa:}\quad & L_\nu\lesssim 10^{29}\,{\rm erg/s/Hz}\quad
 \&\quad T_B\gtrsim 10^{30}\,{\rm K}\,,\label{eq6}\\[1mm]
 &{\rm Type~IIb:}\quad & {\rm otherwise}\,.\label{eq7}
 \eea
 We stress that they are highly speculative, because there are only 17
 repeaters after the robust cut in the first CHIME/FRB catalog
 \cite{CHIMEFRB:2021srp}, which are too few to form a good enough sample
 in statistics. Since the data of repeaters will be rapidly accumulated
 in the future, we hope this subclassification of type~II FRBs could be
 refined. We strongly refer to~\cite{Guo:2022wpf} for the technical
 details of the subclassification.

As mentioned in~\cite{Guo:2022wpf}, there might be three methods to identify
 type~Ib FRBs: (a) Their distribution tracks SFH. But this does not work
 for an individual FRB. (b) The physical criteria similar to
 Eqs.~(\ref{eq4}) and (\ref{eq5}). But this heavily depends on
 cosmology, and a circularity problem might exist. (c) The
 precise localizations of FRBs. We strongly prefer the last
 method (c) in practice. If a non-repeating (type~I)~FRB has been precisely
 localized by using e.g.~VLBI down to the milliarcsecond
 level~\cite{Marcote:2021luf,Marcote:2019sjf} or better in the
 future, its host galaxy and local environment could be well determined.
 Subsequently, it is a type~Ib FRB if this FRB lives in a star-forming
 environment. Of course, its redshift can also be identified accordingly
 in this way. We stress that no circularity problem exists in fact by using
 the method (c). To date, a few dozens of non-repeating FRBs associated
 with young stellar populations (namely type~Ib FRBs) have been already
 determined by the method (c) using telescopes/arrays such as
 DSA-110, ASKAP, Arecibo, Parkes, FAST and EVN, as summarized
 in Table~II of~\cite{Li:2024dge}. Actually, the method (c) ``\,precise
 localizations of FRBs\,'' is the main way to determine type~Ib FRBs
 without circularity problem in practice.

In~\cite{Guo:2022wpf}, we have found that there are some tight empirical
 relations for type~Ia FRBs but not for type~Ib FRBs, and vice versa. These
 make them different in physical properties. Notice that there are only 17
 repeaters after the robust cut in the first CHIME/FRB catalog
 \cite{CHIMEFRB:2021srp}, no empirical relations found for type~II FRBs
 in~\cite{Guo:2022wpf}. On the other hand, there are 65/365 type~Ia/Ib
 FRBs after the robust cut in the first CHIME/FRB catalog
 \cite{CHIMEFRB:2021srp}, respectively. Type~Ib FRBs dominate obviously, and
 hence the empirical relations for them are much tighter than the ones for
 type~Ia FRBs. Unfortunately, the empirical relations only for type~Ia FRBs
 found in~\cite{Guo:2022wpf} do not involve the luminosity distance. In
 addition, type~Ib FRBs are associated with young stellar populations, and
 hence their distribution tracks SFH. This remarkably facilitates generating
 the mock type~Ib FRBs in simulations. Thus, we choose to only calibrate
 type~Ib FRBs as standard candles in the present work.

Actually, we found some tight empirical relations between
 spectral luminosity $L_\nu$\,, isotropic energy $E$ and $\rm DM_E$ for
 type~Ib FRBs in~\cite{Guo:2022wpf}, where $\rm DM_E$ is
 given by Eq.~(\ref{eq3}), and
 \be{eq8}
 L_\nu=4\pi d_L^2 S_\nu\,,\quad\quad E=4\pi d_L^2 \nu_c\, F_\nu/(1+z)\,,
 \ee
 in which $d_L$ is the luminosity distance, $S_\nu$ is the flux, $F_\nu$
 is the specific fluence, $\nu_c$ is the central observing frequency
 ($\nu_c=600\,{\rm MHz}$ for CHIME~\cite{CHIMEFRB:2021srp}). The 2-D
 empirical relations for 365 type~Ib FRBs in the first CHIME/FRB catalog
 are given by~\cite{Guo:2022wpf}
 \bea
 &&\log E=0.8862\log L_\nu+10.0664\,,\label{eq9}\\[1mm]
 &&\log L_\nu=2.4707\log {\rm DM_E}+27.3976\,,\label{eq10}\\[1mm]
 &&\log E=2.2345\log {\rm DM_E}+34.2238\,,\label{eq11}
 \eea
 where ``\,$\log$\,'' gives the logarithm to base 10, and
 $E$, $L_\nu$\,, $\rm DM_E$ are in units of erg, erg/s/Hz, $\dmunit$,
 respectively. In the light of the 2-D empirical relations given by
 Eqs.~(\ref{eq9})\,--\,(\ref{eq11}), it is anticipated that there is a tight
 3-D empirical relation between spectral luminosity $L_\nu$\,, isotropic
 energy $E$ and $\rm DM_E$\,. Fitting to the data, we also found this 3-D
 empirical relation in~\cite{Guo:2022wpf} for 365 type~Ib FRBs in the first
 CHIME/FRB catalog, namely
 \be{eq12}
 \log L_\nu=1.1330\log {\rm DM_E}+0.5986\log E+6.9098\,.
 \ee
 Noting that the empirical relations in Eqs.~(\ref{eq10})\,--\,(\ref{eq12})
 involve $\rm DM_E$\,, they are not suitable for our goal, as mentioned in
 Sec.~\ref{sec1}. Fortunately, the empirical relation in Eq.~(\ref{eq9})
 does not involve DM, but it does involve the luminosity distance $d_L$
 (n.b.~Eq.~(\ref{eq8})), and hence it works well for calibrating type~Ib
 FRBs as standard candles. We strongly refer to~\cite{Guo:2022wpf} for the
 technical details of the empirical relations.

Further, these empirical relations have been
 checked in~\cite{Li:2024dge} with the current $44\sim 52$ localized
 FRBs (most of them are type~Ib FRBs). In~\cite{Guo:2022wpf}, the errors
 were not taken into account. But we have considered the errors by using
 the actual data (rather than simulations) in Sec.~IV of~\cite{Li:2024dge},
 and found that
 \bea
 &&\log E=a\log L_\nu+b\quad {\rm with}\quad
 a=0.8336\pm 0.0375,\ b=11.7858\pm 1.2568, \
 \sigma_{\rm int}=0.3308\,,\label{eq49Liv1}\\[1mm]
 &&\log L_\nu=a\log {\rm DM_E}+b\quad {\rm with}
 \quad a=2.3741\pm 0.4527,\ b=27.8054\pm 1.0893, \
 \sigma_{\rm int}=1.2292\,,\hspace{10mm}\label{eq50Liv1}\\[1mm]
 &&\log E=a\log {\rm DM_E}+b\quad {\rm with}
 \quad a=2.1807\pm 0.3967,\ b=34.4443\pm 0.9568,
 \ \sigma_{\rm int}=1.0591\,,\label{eq51Liv1}
 \eea
 where the constraints on slope $a$ and intercept $b$ are given by their
 means with $1\sigma$ uncertainties. The $1\sigma$, $2\sigma$ and $3\sigma$
 contours for $a$ and $b$ of these empirical relations have also been given
 in Sec.~IV of~\cite{Li:2024dge}. It is reasonable to expect that the errors
 will be significantly decreased in the future with a large amount of
 well-localized type~Ib FRBs.


\section{Calibrating type~Ib FRBs as standard candles}\label{sec3}

It is worth noting that the above empirical relations were found by using
 the first CHIME/FRB catalog~\cite{CHIMEFRB:2021srp}, and the values of
 slopes and intercepts might be changed for the larger and better FRB
 datasets in the future. So, one should not persist in their (exact)
 numerical values. The key point is that these empirical relations do exist
 in such forms for type~Ib FRBs.

Here, we assume that the empirical $L_\nu-E$ relation really exists due
 to the unknown physical mechanism for the origins of type~Ib FRBs. It
 takes the form of Eq.~(\ref{eq9}), namely
 \be{eq13}
 \log\frac{E}{\rm erg}=a\log\frac{L_\nu}{\rm erg/s/Hz}+b\,,
 \ee
 where $a$ and $b$ are both dimensionless constants. In particular,
 $a=0.8862$ and $b=10.0664$~\cite{Guo:2022wpf} for 365 type~Ib FRBs in
 the first CHIME/FRB catalog. Note that the luminosity distance $d_L$
 is implicit in the empirical $L_\nu-E$ relation given by
 Eq.~(\ref{eq13}). Using Eq.~(\ref{eq8}), we recast Eq.~(\ref{eq13}) as
 \be{eq14}
 \mu=-\frac{5}{2\left(1-\alpha\right)}\,\log\frac{\;F_\nu/(1+z)\,}{\rm
 Jy\;ms}+\frac{5\alpha}{2\left(1-\alpha\right)}\,\log\frac{S_\nu}{\;\rm
 Jy\,}+\beta\,,
 \ee
 where $\alpha=a\not=1$, $\beta={\rm const.}$ is a complicated combination
 of $a$, $b$, $\nu_c/{\rm MHz}$, and $\mu$ is the well-known distance
 modulus defined by
 \be{eq15}
 \mu=5\log\frac{d_L}{\,\rm Mpc}+25\,.
 \ee
 Unlike $L_\nu$ and $E$ in Eq.~(\ref{eq13}), we note that $F_\nu$\,, $S_\nu$
 and $\mu$ (equivalently $d_L$) are all observed quantities, and hence
 it is more convenient to instead use Eq.~(\ref{eq14}) in calibrating
 type~Ib FRBs as standard candles. There are two free parameters $\alpha$
 and $\beta$ in Eq.~(\ref{eq14}), and they will be calibrated by using
 type~Ib FRBs at low redshifts.

Of course, current data of FRBs are certainly not enough to calibrate
 type~Ib FRBs as standard candles.~So, it will be a proof of concept
 by using the simulated FRBs in the present work.~We assume that there
 will be a large amount of type~Ib FRBs with identified redshifts in the
 future. If a non-repeating (type~I) FRB has been precisely localized
 by using e.g.~VLBI down to the milliarcsecond level~\cite{Marcote:2021luf,
 Marcote:2019sjf} or better in the future, its host galaxy and local
 environment can be well determined. Subsequently, it is a type~Ib FRB
 if this FRB lives in a star-forming environment. Its redshift could
 be identified by using the spectra of the host galaxy.~Actually, as
 summarized in Table~II of~\cite{Li:2024dge}, 52 localized FRBs (most of
 them are type~Ib FRBs) were found recently by using telescopes/arrays
 such as DSA-110, ASKAP, Arecibo, Parkes, FAST and EVN. After
 Ref.~\cite{Li:2024dge} has been submitted to arXiv, another 48 new
 localized FRBs (most of them are type~Ib FRBs) were released
 in e.g.~\cite{Tian:2024ygd,Sharma:2024fsq,Connor:2024mjg}. Note that
 almost all these $\sim 100$ localized FRBs (most of them are type~Ib FRBs)
 were observed and localized in the recent $2\sim 3$ years. Actually,
 many new and powerful projects to precisely localize FRBs have been
 proposed and under construction, such as DSA-2000, FASTA, CHIME/FRB
 outriggers and EVN. Therefore, it is reasonable to expect that in the near
 future ($3\sim 5$ years) there will be a large amount ($300\sim 500$)
 of type~Ib FRBs with identified redshifts, and the number might be
 $\sim {\cal O}(10^3)$ in the next 10 years.

On the other hand, the luminosity distance $d_L$ (equivalently the distance
 modulus $\mu$) of type~Ib FRBs should be independently measured to build
 the Hubble diagram. This is a fairly difficult task at high redshifts. But
 it is possible to measure the luminosity distance of type~Ib FRBs at enough
 low redshifts. Fortunately, the redshift range of FRBs is very wide. They
 could be at very high redshifts $z>3$ (even $z\sim 15$
 \cite{Zhang:2018csb}), and hence they might be a powerful probe for the
 early universe. On the other hand, they can also be at very low redshifts,
 even in our Milky Way ($z=0$) as shown by the Galactic FRB 200428. In fact,
 many nearby FRBs at very low identified redshifts were found
 \cite{Heintz:2020}. Thus, it is possible that a type~Ib FRB and a SNIa or
 a Cepheid are in the same host galaxy, and hence the luminosity distance
 $d_L$ (equivalently the distance modulus $\mu$) of type~Ib FRB is equal to
 the one of SNIa or Cepheid. Note that type~Ib FRB and SNIa/Cepheid are not
 necessarily associated. This is just similar to the case of calibrating
 SNIa as standard candles by using Cepheids, as mentioned in
 Sec.~\ref{sec1}. Because Cepheids are certainly at very low redshifts and
 SNIa can be at not so low redshifts (for example, $z\sim 0.1$ or $0.5$), we
 only consider the case of SNIa in this work. As is well known, currently
 the distance modulus $\mu$ of SNIa can be measured very precisely. It is
 also expected that the precision will be significantly improved in the
 future. Thus, it is reasonable to assume that the distance modulus $\mu$
 of type~Ib FRB at low redshifts could also be measured in the future with
 the precision comparable with the one of SNIa.

If the number of type~Ib FRBs at low redshifts with measured distance moduli
 $\mu$, redshifts $z$, fluence $F_\nu$ and flux $S_\nu$ is large enough, the
 empirical relation in Eq.~(\ref{eq14}) could be well calibrated by fitting
 it to these type~Ib FRBs (note that ``\,how many type~Ib FRBs are large
 enough\,'' is the object to be studied in Secs.~\ref{sec4b}
 and \ref{sec4c}, and we will soon find that $N_{\rm FRB}=300$ or $500$ are
 acceptable, as claimed at the end of Sec.~\ref{sec4c}). In this way,
 the free parameters $\alpha$ and $\beta$ in Eq.~(\ref{eq14}) and their
 uncertainties will be determined.~Then, we assume this calibrated empirical
 relation in Eq.~(\ref{eq14}) is universal, namely it also holds at high
 redshifts (note that the well-known empirical Phillips relation for
 SNIa~\cite{Phillips:1993ng,Riess:2021jrx} and the well-known empirical
 Amati relation for GRBs~\cite{Liang:2008kx,Wei:2010wu,Amati:2002ny,
 Schaefer:2006pa,Wei:2008kq,Liu:2014vda}) are also assumed to
 be universal due to the (unknown) physical mechanisms (in the case of GRBs
 it is still unknown to date), but their successful applications in
 cosmology justified these assumptions. Similarly, it is better to keep
 an open mind to the universalness of the empirical relation
 for FRBs.~In principle, one might test this empirical relation in different
 redshift ranges and see whether its parameters are consistent (we thank the
 referee for pointing out this issue), like in the cases of SNIa and
 GRBs.~But currently there are not enough actual data to do this, and we
 hope it could be done in the next 5 years). Now, for each type~Ib FRB
 at high redshift, its redshift $z$, fluence $F_\nu$ and flux $S_\nu$
 have been observed, and the parameters $\alpha$, $\beta$ take the same
 values determined at low redshift in the previous step. Consequently,
 its distance modulus $\mu$ (equivalently the luminosity distance $d_L$) in
 the left hand side of Eq.~(\ref{eq14}) can be derived since the quantities
 in the right hand side of Eq.~(\ref{eq14}) are all known now. Its error
 $\sigma_\mu$ can also be obtained by using the error propagation. So,
 the Hubble diagram of type~Ib FRBs at high redshifts is on hand.~We
 can use these type~Ib FRBs with known distance moduli $\mu$, errors
 $\sigma_\mu$, and redshifts $z$ to constrain the cosmological models,
 similar to the case of SNIa.


\section{Key factors affecting
 the calibration and the~cosmological~constraints}\label{sec4}

As mentioned above, current data of FRBs are not enough to calibrate
 type~Ib FRBs as standard candles. Thus, we have to use the mock type~Ib
 FRBs instead. But the same pipeline holds for the actual type~Ib FRBs in
 the future.~In the followings, we briefly describe how to generate the
 mock type~Ib FRBs, and calibrate type~Ib FRBs as standard candles.~Then,
 we use the calibrated type~Ib FRBs at high redshifts to constrain the
 cosmological model.~Clearly, there are many key factors affecting the
 calibration and the cosmological constraints. We will test them one by one.


\subsection{Generating the mock type~Ib FRBs}\label{sec4a}

Here, we generate the mock type~Ib FRBs closely following the method
 used in~\cite{Qiang:2021ljr,Guo:2022wpf} (see also~\cite{Zhang:2021kdu}),
 but the distribution of type~Ib FRBs should track SFH. The
 mock observed type~Ib FRB redshift rate distribution is
 given by~\cite{Qiang:2021ljr,Guo:2022wpf,Zhang:2021kdu}
 \be{eq16}
 \frac{dN}{dt_{\rm obs}\,dz}=\frac{1}{1+z}\cdot\frac{dN}{dt\,dV}\cdot
 \frac{c}{H_0}\cdot\frac{4\pi d_C^2}{h(z)}\,,
 \ee
 where the comoving distance $d_C=d_L/(1+z)$\,, and $dt/dt_{\rm obs}=
 (1+z)^{-1}$ due to the cosmic expansion is used. In the present work, we
 consider the flat $\Lambda$CDM cosmology, and hence the dimensionless
 Hubble parameter $h(z)$ and the luminosity distance $d_L$ read
 \be{eq17}
 h(z)\equiv H(z)/H_0=\left[\,\Omega_m\left(1+z\right)^3+\left(1-\Omega_m
 \right)\,\right]^{1/2}\,,\quad\quad
 d_L=\left(1+z\right)\frac{c}{H_0}\int_0^z\frac{d\tilde{z}}{h(\tilde{z})}\,.
 \ee
 Here, we adopt $\Omega_m=0.3153$ and $H_0=67.36\;{\rm km/s/Mpc}$ from the
 Planck 2018 results~\cite{Aghanim:2018eyx}.~Since the intrinsic type~Ib
 FRB redshift distribution tracks SFH, we have $dN/(dt\,dV)\propto {\rm
 SFH}(z)$, while the latest result from the observations for SFH is given
 by~\cite{Madau:2016jbv} (see also~\cite{Qiang:2021ljr,Guo:2022wpf})
 \be{eq18}
 {\rm SFH}(z)\propto\frac{(1+z)^{2.6}}{1+\left((1+z)/3.2\right)^{6.2}}\,.
 \ee
 On the other hand, we generate the isotropic energy $E$ for
 the mock type~Ib FRBs with~\cite{Qiang:2021ljr,Guo:2022wpf,Zhang:2021kdu}
 \be{eq19}
 dN/dE\propto\left(E/E_c\right)^{-s}\exp\left(-E/E_c\right)\,,
 \ee
 where we adopt $s=1.9$ and $\log\left(E_c/{\rm erg}\right)=41$, well
 consistent with the observations~\cite{Qiang:2021ljr,Guo:2022wpf}. We
 generate $N_{\rm sim}$ mock type~Ib FRBs as follows: (i) for each mock
 type~Ib FRB, randomly assign a mock redshift $z$ to this FRB from the
 redshift distribution in Eq.~(\ref{eq16}); (ii) generate a mock energy
 $E_{\rm int}$ randomly from the distribution in Eq.~(\ref{eq19}) for
 this FRB; (iii) derive the luminosity distance $d_{L\cmn {\rm int}}$ and
 the distance modulus $\mu_{\rm int}$ by using Eqs.~(\ref{eq17}) and
 (\ref{eq15}) with the mock redshift $z$ for this FRB; (iv) derive the
 fluence $F_{\nu\cmn {\rm int}}$ by using Eq.~(\ref{eq8}) with
 $E_{\rm int}$, $d_{L\cmn {\rm int}}$, $\nu_c=600\,{\rm MHz}$ (the one of
 CHIME~\cite{CHIMEFRB:2021srp}) and the mock redshift $z$ for this FRB;
 (v) assign an error $\sigma_{F\cmn {\rm obs}}=\sigma_{F\cmn {\rm rel}}
 \,F_{\nu\cmn {\rm int}}$ to the ``\,observed\,'' fluence
 $F_{\nu\cmn {\rm obs}}$ for this FRB, while $F_{\nu\cmn {\rm obs}}$ is
 randomly assigned from a Gaussian distribution with the mean $F_{\nu\cmn
 {\rm int}}$ and the standard deviation $\sigma_{F\cmn {\rm obs}}$. Note
 that the relative error $\sigma_{F\cmn {\rm rel}}$ will be specified below;
 (vi) repeat the above steps for $N_{\rm sim}$ times. Finally, $N_{\rm sim}$
 mock type~Ib FRBs are on hand.


 \begin{center}
 \begin{figure}[tb]
 \centering
 \vspace{-8mm} \hspace{-6mm} 
 \includegraphics[width=0.48\textwidth]{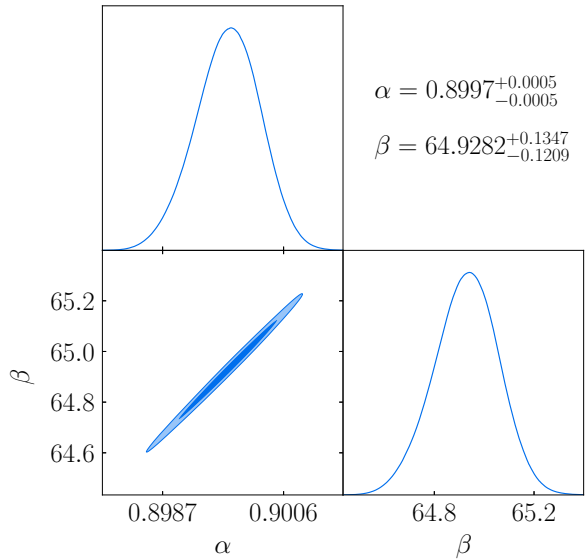}
 \vspace{-2mm}  
 \caption{\label{fig1} The marginalized $1\sigma$ constraints on the
 parameters $\alpha$, $\beta$ in the empirical relation~(\ref{eq14}) and
 their contours from the data of type~Ib FRBs at low redshifts $z<z_d$
 for the fiducial case. Note that in order to better quantify
 the uncertainties on the parameters $\alpha$ and $\beta$, the fit would
 need to be performed again after centering the data. See
 Sec.~\ref{sec4b} for details.}
 \end{figure}
 \end{center}


\vspace{-10.9mm} 

But these $N_{\rm sim}$ mock type~Ib FRBs intrinsically generated above are
 not the ones ``\,detected\,'' by the telescope, due to the telescope's
 sensitivity threshold and instrumental selection effects near
 the threshold. So, the next step is to filter them by using the telescope's
 sensitivity model, which is chosen to be the one for CHIME considered
 in~\cite{Qiang:2021ljr,Guo:2022wpf,Zhang:2021kdu}. The sensitivity
 threshold is $\log F_{\nu\cmn {\rm min}}=-0.5$ for CHIME, where the
 specific fluence is in units of Jy\,ms. The FRBs with fluences below this
 threshold cannot be detected. On the other hand, there is a ``\,gray
 zone\,'' in the $\log F_\nu$ distribution, within which CHIME has not
 reached full sensitivity to all sources, due to the direction-dependent
 sensitivity of the telescope~\cite{Qiang:2021ljr,Guo:2022wpf,
 Zhang:2021kdu}. The detection efficiency parameter in the ``\,gray zone\,''
 is given by $\eta_{\rm det}={\cal R}^3$, where ${\cal R}=(\log F_{\nu\cmn
 {\rm th}}-\log F_{\nu\cmn {\rm th}}^{\rm min})/(\log F_{\nu\cmn {\rm
 th}}^{\rm max}-\log F_{\nu\cmn {\rm th}}^{\rm min})$\,, such
 that $\eta_{\rm det}\to 0$ at $\log F_{\nu\cmn {\rm th}}^{\rm min}=-0.5$
 and $\eta_{\rm det}\to 1$ at $\log F_{\nu\cmn {\rm th}}^{\rm max}$. Outside
 the ``\,gray zone\,'', $\eta_{\rm det}=1$. For type~Ib FRBs tracking SFH,
 we set $\log F_{\nu\cmn {\rm th}}^{\rm max}=0.9$, which is very close to
 the best value found in Table~II of~\cite{Guo:2022wpf}.

In the present work, we generate a pool of mock type~Ib FRBs
 ``\,detected\,'' by the telescope to save the computational power and time.
 That is, we randomly generate $N_{\rm sim}=500,000,000$ mock type~Ib FRBs,
 and then filter them by using the telescope's sensitivity model, as
 mentioned above. Finally, about 40,000 mock ``\,detected\,'' type~Ib FRBs
 after the filter enter this pool. When $N_{\rm FRB}$ mock type~Ib FRBs
 are needed in the following subsections, we randomly sample them from
 this pool.

The next step is to assign the mock flux and its error for each mock type~Ib
 FRB in this pool.~We assume that the underlying empirical relation in
 Eq.~(\ref{eq14}) is actually
 \be{eq20}
 \mu=-25\log\frac{\;F_\nu/(1+z)\,}{\rm Jy\;ms}+22.5\log\frac{S_\nu}
 {\;\rm Jy\,}+65\,,
 \ee
 which corresponds to $\alpha=0.9$ and $\beta=65$, or equivalently $a=0.9$
 and $b\simeq 10.1$ approximately in Eq.~(\ref{eq13}). We can derive $S_{\nu
 \cmn {\rm int}}$ by using Eq.~(\ref{eq20}) with $\mu_{\rm int}$, $F_{\nu
 \cmn {\rm int}}$ and the mock redshift $z$ for this FRB. We assign an error
 $\sigma_{S\cmn {\rm obs}}=\sigma_{S\cmn {\rm rel}}\,S_{\nu\cmn {\rm int}}$
 to the ``\,observed\,'' flux $S_{\nu\cmn {\rm obs}}$ for this FRB, while
 $S_{\nu\cmn {\rm obs}}$ is randomly assigned from a Gaussian distribution
 with the mean $S_{\nu\cmn {\rm int}}$ and the standard deviation $\sigma_{S
 \cmn {\rm obs}}$. Similarly, we assign an error $\sigma_{\mu\cmn {\rm obs}}
 =\sigma_{\mu\cmn {\rm rel}}\,\mu_{\rm int}$ to the ``\,observed\,''
 distance modulus $\mu_{\rm obs}$ for this FRB, while $\mu_{\rm obs}$ is
 randomly assigned from a Gaussian distribution with the mean $\mu_{\rm
 int}$ and the standard deviation $\sigma_{\mu\cmn {\rm obs}}$. Note that
 the relative errors $\sigma_{S\cmn {\rm rel}}$ and $\sigma_{\mu\cmn {\rm
 rel}}$ will be specified below.

So far, a pool of about 40,000 mock ``\,detected\,'' type~Ib FRBs with the
 ``\,observed\,'' redshifts $z$, fluences $F_{\nu\cmn {\rm obs}}$ and their
 errors $\sigma_{F\cmn {\rm obs}}$, fluxes $S_{\nu\cmn {\rm obs}}$ and
 their errors $\sigma_{S\cmn {\rm obs}}$, are ready. In addition, the
 ``\,observed\,'' distance moduli $\mu_{\rm obs}$ and their
 errors $\sigma_{\mu\cmn {\rm obs}}$ for the mock type~Ib FRBs at low
 redshits are also on hand. We will pretend to use them as the actual
 ones blindly in the followings.


 \begin{center}
 \begin{figure}[tb]
 \centering
 \vspace{-8mm} \hspace{-6mm} 
 \includegraphics[width=0.415\textwidth]{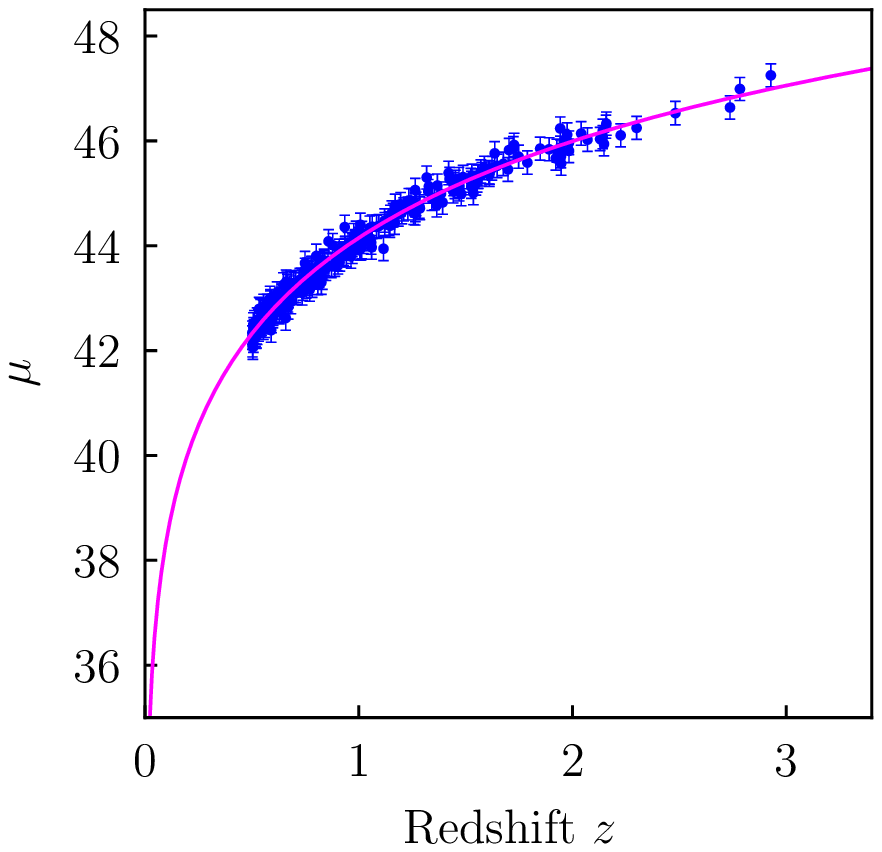}\quad
 \includegraphics[width=0.36\textwidth]{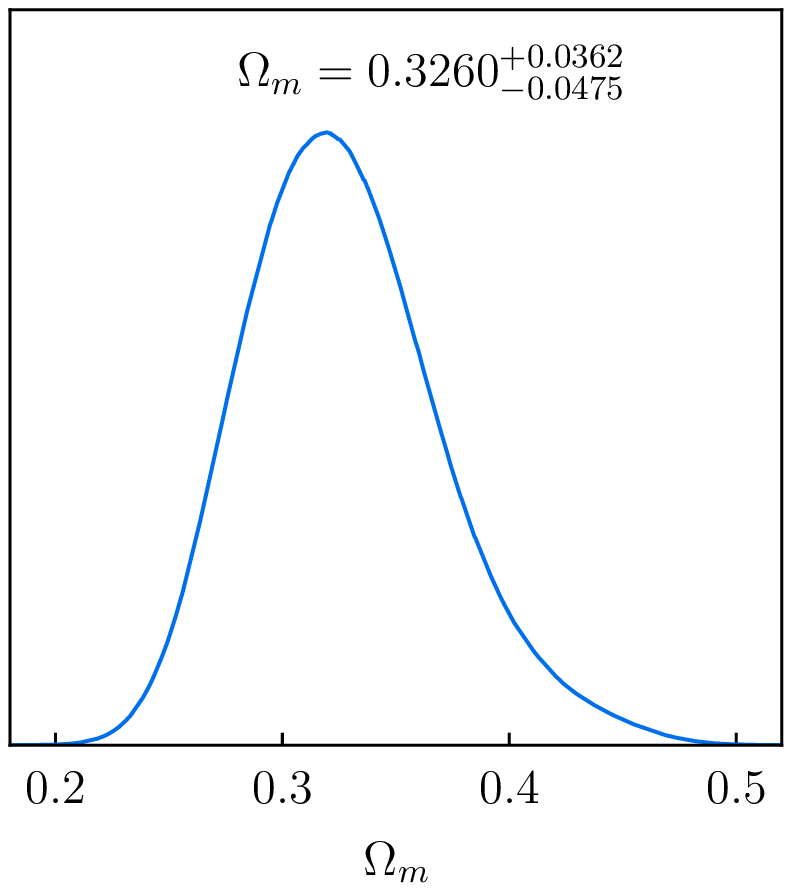}
 \vspace{-2mm}  
 \caption{\label{fig2} Left panel: The Hubble diagram $\mu$ versus $z$ for
 type~Ib FRBs at high redshifts $z\geq z_d$\,. The magenta line indicates
 the flat $\Lambda$CDM cosmology with the best-fit $\Omega_m$ given in the
 right panel. Right panel: The marginalized $1\sigma$ constraint on the
 parameter $\Omega_m$ in the flat $\Lambda$CDM cosmology from the data of
 type~Ib FRBs at high redshifts $z\geq z_d$\,. Note that they are both
 for the fiducial case. See Sec.~\ref{sec4b} for details.}
 \end{figure}
 \end{center}


\vspace{-9.5mm} 


\subsection{The fiducial case}\label{sec4b}

At first, we consider the fiducial case, and the other cases below will
 be compared with it.~In this fiducial case, we consider
 $N_{\rm FRB}=500$ mock type~Ib FRBs, which can be randomly sampled from
 the pool built in Sec.~\ref{sec4a} (note that in principle we could
 alternatively generate 500 mock type~Ib FRBs one by one following the
 instructions in Sec.~\ref{sec4a}, but there is no difference between
 this way and randomly sampling from the pool. Actually, the
 mock $F_\nu$ and $S_\nu$ are not directly generated. They come from the
 mock $d_{L\cmn {\rm int}}$ and $E_{\rm int}$ (which is randomly sampled
 from the distribution in Eq.~(\ref{eq19}) with unequal chance), and
 they have been filtered by using the telescope's sensitivity model (namely
 FRBs are not detected with the same chance), as described in
 Sec.~\ref{sec4a}. Actually, this is a standard pipeline extensively used in
 the literature).~On the other hand, although the fluence and flux of
 FRBs cannot be measured with high precision currently by the telescope
 (e.g.~CHIME), we assume that they could be measured precisely in the
 future. So, in the fiducial case, we set the relative errors $\sigma_{F\cmn
 {\rm rel}}$ and $\sigma_{S\cmn {\rm rel}}$ to be both $1\%$ (frankly,
 we cannot give the timetable to achieve this goal. But many new and
 giant telescopes are proposed or under construction around the world,
 especially the promising Square Kilometre Array (SKA) in Australia and
 South Africa, FAST Array (FASTA) in China, while FRB detection has
 been one of the main scientific goals in the ambitious visions of many
 countries. The era of FRBs is coming in the near future. Let us be
 optimistical and keep an open mind). As mentioned above, we obtain the
 distance modulus of type~Ib FRB at low redshift by equaling it to the
 one of SNIa in the same host galaxy. As is well known, currently the
 distance modulus $\mu$ of SNIa can be measured very precisely. It is
 expected that the precision will be significantly improved in
 the future. For example, the expected aggregate precision of SNIa detected
 by the Roman Space Telescope (formerly WFIRST, planned for launch in
 the mid-2020s) is $0.2\%$ at $z<1$~\cite{Spergel:2013}. So, in the fiducial
 case, we set the relative error $\sigma_{\mu\cmn {\rm rel}}=0.2\%$ for
 type~Ib FRBs at low redshifts. Of course, we need a redshift divide
 $z_d$ to define ``\,low\,'' ($z<z_d$) and ``\,high\,'' ($z\geq
 z_d$) redshifts. In the fiducial case, we set $z_d=0.5$.


 \begin{center}
 \begin{figure}[tb]
 \centering
 \vspace{-8mm} \hspace{-6mm} 
 \includegraphics[width=0.83\textwidth]{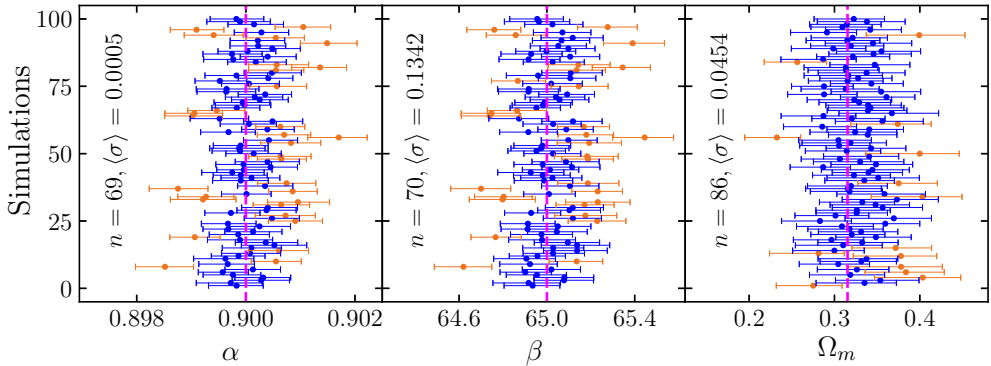}
 \vspace{-2mm} 
 \caption{\label{fig3} The marginalized $1\sigma$ constraints on the
 parameters $\alpha$ (left panel), $\beta$ (middle panel) in the empirical
 relation~(\ref{eq14}) and the cosmological parameter $\Omega_m$ (right
 panel) for 100 simulations in the fiducial case. The blue means with error
 bars (the chocolate means with error bars) indicate that the assumed
 $\alpha=0.9$, $\beta=65$ or $\Omega_m=0.3153$ (indicated by the magenta
 dashed lines) are consistent (inconsistent) with the mock type~Ib FRBs
 within $1\sigma$ region, respectively. In each panel, $n$ and $100-n$ are
 the numbers of blue and chocolate means with error bars, respectively.
 $\langle\sigma\rangle$ is the mean of the uncertainties of 100 constraints
 on the parameters $\alpha$ (left panel), $\beta$ (middle panel) and
 $\Omega_m$ (right panel), respectively. See Sec.~\ref{sec4b} for details.}
 \end{figure}
 \end{center}


\vspace{-10.8mm} 

Following the instructions in Sec.~\ref{sec3}, we calibrate the
 empirical relation~(\ref{eq14}) by using type~Ib FRBs at low redshifts
 $z<z_d$ (note that directly fitting the empirical relation to the data
 at low redshifts is the way extensively used in the similar field of
 calibrating long GRBs as standard candles (see e.g.~\cite{Liang:2008kx,
 Wei:2010wu}). It is expected that the correlation between $\alpha$ and
 $\beta$ would diminish significantly if the mean values of $F_\nu$ and
 $S_\nu$ are subtracted from each object, namely centering the data (we
 thank the referee for pointing out this issue). One might try this way
 alternatively, while we leave it here as an open topic). To this end,
 we use the Markov Chain Monte Carlo (MCMC) code
 Cobaya~\cite{Torrado:2020dgo}, which is the Python version of
 the well-known CosmoMC~\cite{Lewis:2002ah}. Fitting the empirical relation
 given by Eq.~(\ref{eq14}) to the data of type~Ib FRBs at low redshifts
 $z<z_d$\,, we find the marginalized $1\sigma$ constraints on the parameters
 $\alpha$ and $\beta$, namely
 \be{eq21}
 \alpha=0.8997^{+0.0005}_{+0.0005}\,,\quad\quad
 \beta=64.9282^{+0.1347}_{-0.1209}\,,
 \ee
 and we also present the contours and the marginalized probability in
 Fig.~\ref{fig1}. Then, assuming the calibrated empirical
 relation~(\ref{eq14}) with $\alpha$ and $\beta$ given in Eq.~(\ref{eq21})
 still holds at high redshifts $z\geq z_d$\,, we can derive the distance
 moduli $\mu$ for type~Ib FRBs at high redshifts $z\geq z_d$ with their
 observed fluences $F_{\nu\cmn {\rm obs}}$ and fluxes $S_{\nu\cmn {\rm
 obs}}$. Their errors $\sigma_\mu$ can also be obtained by using the error
 propagation. In the left panel of Fig.~\ref{fig2}, we present the Hubble
 diagram $\mu$ versus $z$ for type~Ib FRBs at high redshifts $z\geq z_d$\,,
 while the error bars $\sigma_\mu$ are also plotted. Fitting the flat
 $\Lambda$CDM cosmology given by Eq.~(\ref{eq17}) to the distance moduli
 $\mu(z_i)$ data of type~Ib FRBs at high redshifts $z\geq z_d$\,, we obtain
 the $1\sigma$ constraint on $\Omega_m$, namely
 \be{eq22}
 \Omega_m=0.3260^{+0.0362}_{-0.0475}\,,
 \ee
 while the Hubble constant $H_0$ has been marginalized.~We also
 present the marginalized probability in the right panel of
 Fig.~\ref{fig2}.~Clearly, the assumed value $\Omega_m=0.3153$ used in
 Sec.~\ref{sec4a} to generate the mock type~Ib FRBs is well consistent with
 the one in Eq.~(\ref{eq22}).


 \begin{center}
 \begin{figure}[tb]
 \centering
 \vspace{-9mm} \hspace{-6mm} 
 \includegraphics[width=0.82\textwidth]{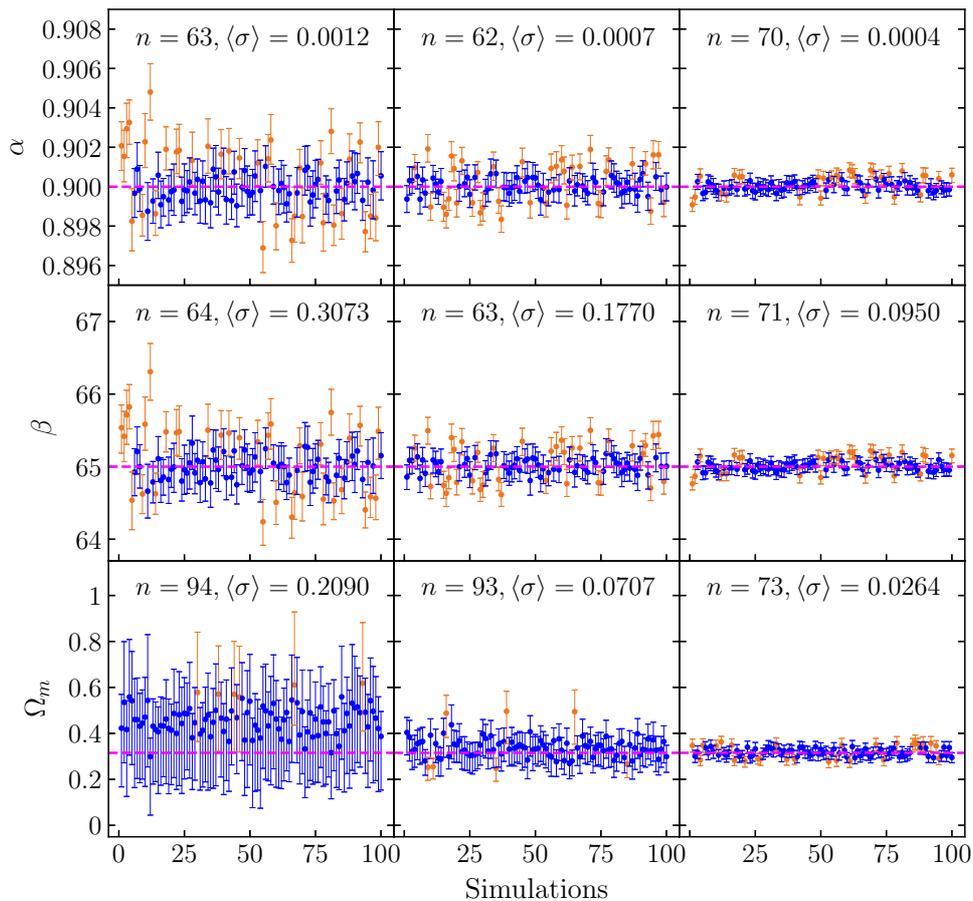}
 \vspace{-2mm} 
 \caption{\label{fig4} The same as in Fig.~\ref{fig3}, but for the cases
 of $N_{\rm FRB}=100$ (left panels), 300 (middle panels) and 1000 (right
 panels) type~Ib FRBs, respectively. Fig.~\ref{fig3} ($N_{\rm FRB}=500$)
 should be viewed together. See Sec.~\ref{sec4c} for details.}
 \end{figure}
 \end{center}



 \begin{center}
 \begin{figure}[tb]
 \centering
 \vspace{-9mm} \hspace{-6mm} 
 \includegraphics[width=0.82\textwidth]{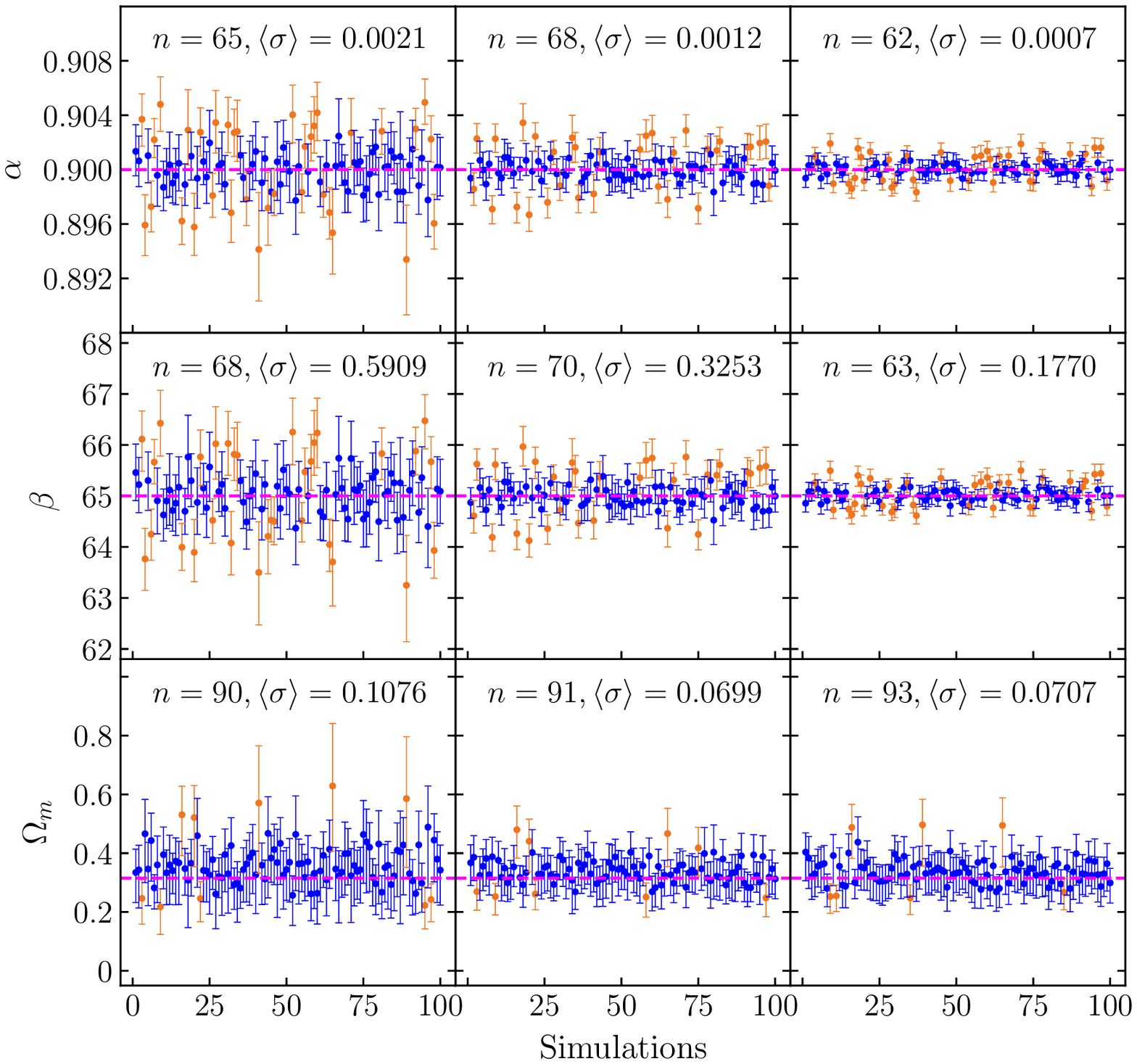}
 \vspace{-2mm} 
 \caption{\label{fig5} The same as in Fig.~\ref{fig3}, but for the cases
 of $N_{\rm FRB}=300$ and $z_d=0.1$ (left panels), 0.2 (middle panels) and
 0.5 (right panels), respectively. See Sec.~\ref{sec4d} for details.}
 \end{figure}
 \end{center}



 \begin{center}
 \begin{figure}[tb]
 \centering
 \vspace{-7mm} \hspace{-6mm} 
 \includegraphics[width=0.82\textwidth]{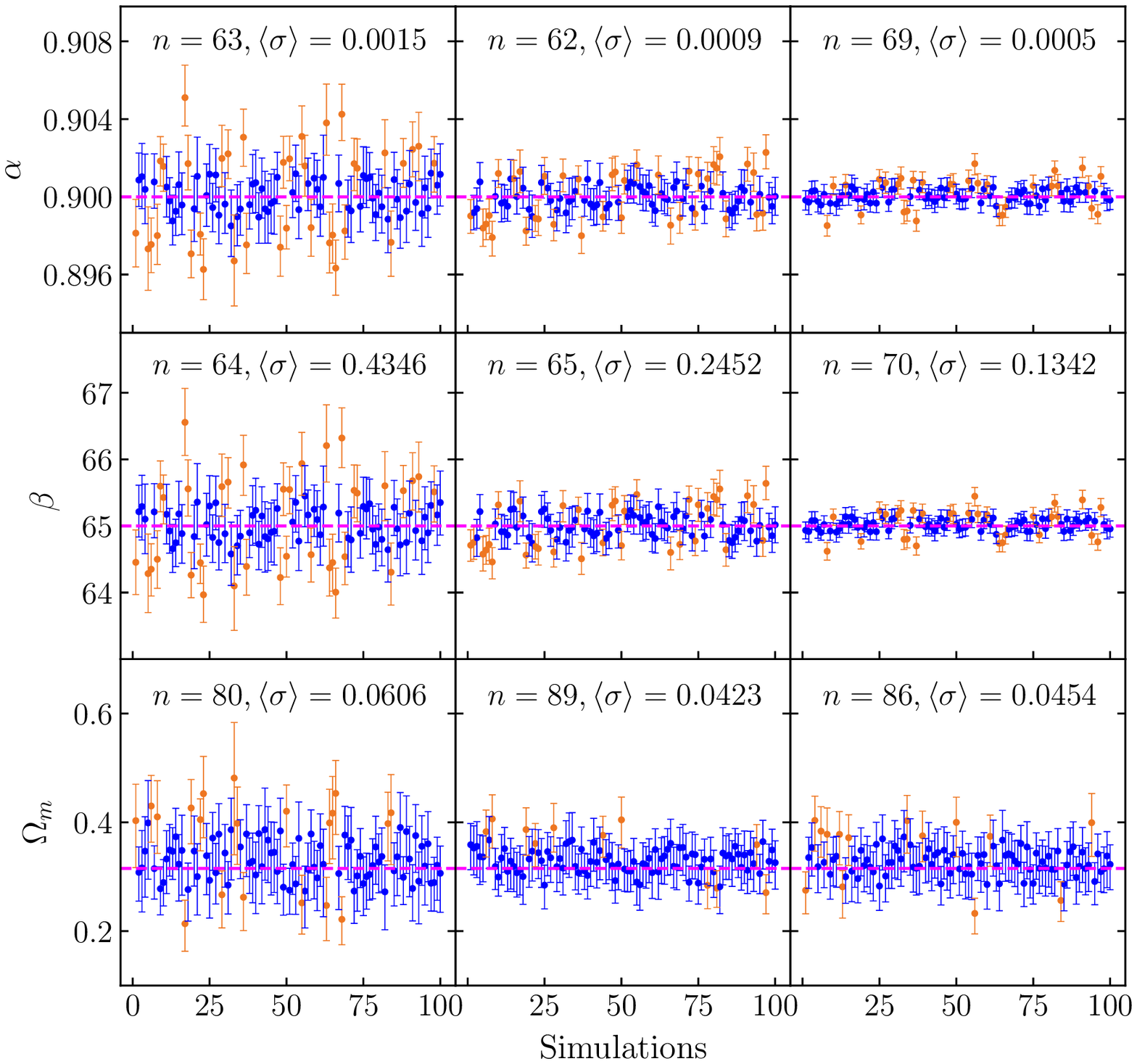}
 \vspace{-2mm} 
 \caption{\label{fig6} The same as in Fig.~\ref{fig3}, but for the cases
 of $N_{\rm FRB}=500$ and $z_d=0.1$ (left panels), 0.2 (middle panels) and
 0.5 (right panels), respectively. See Sec.~\ref{sec4d} for details.}
 \end{figure}
 \end{center}


\vspace{-20mm} 

To avoid the statistical noise due to random fluctuations, one should
 repeat the above constraints for a large number of simulations.~But it
 is expensive to consider too many simulations since they consume a
 large amount of computation power and time.~As a balance, here we consider
 100 simulations, which is enough actually (in principle, the number is
 ``\,large enough\,'' to repeat at least ${\cal O}(10^6)$ simulations,
 but it is too expensive for us.~On the other hand, there is no significant
 difference between ${\cal O}(10^2)$ and ${\cal O}(10^3)$ simulations.~One
 might try ${\cal O}(10^4)$ simulations to this end, but we have to only
 consider the affordable 100 simulations in the present work to save our
 poor computational resource). In Fig.~\ref{fig3}, we present
 the marginalized $1\sigma$ constraints on the parameters $\alpha$, $\beta$
 in the empirical relation~(\ref{eq14}) and the cosmological parameter
 $\Omega_m$ for 100 simulations in the fiducial case. It is easy to see
 from Fig.~\ref{fig3} that the assumed values $\alpha=0.9$, $\beta=65$, and
 $\Omega_m=0.3153$ used in Sec.~\ref{sec4a} to generate the mock type~Ib
 FRBs can be found within $1\sigma$ region in most of the 100
 simulations ($\sim 70\%$ for $\alpha$ and $\beta$, $86\%$ for $\Omega_m$).
 This implies that the constraints from these mock type~Ib FRBs are
 fairly reliable and robust. The pipeline of calibrating type~Ib FRBs as
 standard candles to study cosmology works well.~In the followings, we will
 test the key factors affecting the calibration and the cosmological
 constraints, by considering other cases different from the fiducial case.

\vspace{-2.8mm} 


\subsection{The number of type~Ib FRBs}\label{sec4c}

In the fiducial case, we have set $N_{\rm FRB}=500$, $z_d=0.5$,
 $\sigma_{\mu\cmn {\rm rel}}=0.2\%$, and $\sigma_{F\cmn {\rm rel}}=
 \sigma_{S\cmn {\rm rel}}=1\%$. Clearly, the number of type~Ib FRBs (namely
 $N_{\rm FRB}$) plays an important role.~It is easy to expect that the
 constraints become better for the larger $N_{\rm FRB}$.~However, the
 time we have to wait becomes longer to accumulate a larger amount of
 type~Ib FRBs with identified redshifts.~So, the suitable question is
 how many type~Ib FRBs with identified redshifts are enough to
 get the acceptable constraints?

Here, we modify the fiducial case by considering $N_{\rm FRB}=100$, 300,
 and 1000, while the other settings keep unchanged. Following
 the similar pipeline in Sec.~\ref{sec4b} for the fiducial case, in
 Fig.~\ref{fig4} we present the marginalized $1\sigma$ constraints on
 the parameters $\alpha$, $\beta$ in the empirical relation~(\ref{eq14})
 and the cosmological parameter $\Omega_m$ for 100 simulations in the
 cases of $N_{\rm FRB}=100$, 300, and 1000. Note that Fig.~\ref{fig4}
 should be viewed together with Fig.~\ref{fig3} ($N_{\rm FRB}=500$). It
 is easy to see from Figs.~\ref{fig3} and~{\ref{fig4} that the assumed
 values $\alpha=0.9$, $\beta=65$, and $\Omega_m=0.3153$ used in
 Sec.~\ref{sec4a} to generate the mock type~Ib FRBs can be found within
 $1\sigma$ region in most of the 100 simulations (namely $62\sim 71\%$
 for $\alpha$ and $\beta$, $73\sim 94\%$ for $\Omega_m$). This implies
 that the constraints from these mock type~Ib FRBs are fairly reliable
 and robust.~Clearly, we find from Figs.~\ref{fig3} and~{\ref{fig4} that
 the constraints on $\alpha$, $\beta$ and $\Omega_m$ become better for
 larger $N_{\rm FRB}$, as expected. In particular, the mean of
 the uncertainties $\langle\sigma\rangle=0.2090$ for $\Omega_m$ in the case
 of $N_{\rm FRB}=100$ (bottom-left panel of Fig.~\ref{fig4}) is too
 large ($\sim 66\%$) compared with the assumed value $\Omega_m=0.3153$. So,
 $N_{\rm FRB}=100$ is not enough to get the acceptable constraints.~But the
 cases of $N_{\rm FRB}=300$ and 500 are acceptable, since the uncertainties
 decrease dramatically.~Although $N_{\rm FRB}=1000$ is best, the cost is
 expensive to accumulate such a large amount of type~Ib FRBs with identified
 redshifts (note that as mentioned in Sec.~\ref{sec3} (namely in the
 next paragraph below Eq.~(\ref{eq15})), we estimate that $N_{\rm FRB}$
 could be $\sim {\cal O}(10^3)$ in the next 10 years with the future
 giant telescopes/arrays such as SKA and FASTA). So, we only consider
 the cases of $N_{\rm FRB}=300$ and 500 in the followings.


\subsection{The redshift divide}\label{sec4d}

It is expected that the closer FRBs are easier to be detected and localized
 in the host galaxies. So, the lower redshift divide $z_d$ might be better.
 But it cannot be very low, otherwise there will be not enough type~Ib FRBs
 at $z<z_d$ to calibrate the empirical relation~(\ref{eq14}) with an
 acceptable precision. We try to find a suitable redshift divide $z_d$
 in the balance.

We consider the redshift divides $z_d=0.1$, 0.2 and 0.5 for the cases of
 $N_{\rm FRB}=300$ and 500, respectively, while the other settings of
 the fiducial case keep unchanged.~Following the similar pipeline in
 Sec.~\ref{sec4b} for the fiducial case, in Figs.~\ref{fig5} and~\ref{fig6}
 we present the marginalized $1\sigma$ constraints on the parameters
 $\alpha$, $\beta$ and $\Omega_m$ for 100 simulations in the cases of
 $N_{\rm FRB}=300$ and 500, respectively.~We find from Figs.~\ref{fig5}
 and~\ref{fig6} that the constraints from the mock type~Ib FRBs are fairly
 reliable and robust, since the assumed values $\alpha=0.9$, $\beta=65$, and
 $\Omega_m=0.3153$ used in Sec.~\ref{sec4a} to generate the mock type~Ib
 FRBs can be found within $1\sigma$ region in most of the 100 simulations
 (namely $62\sim 70\%$ for $\alpha$ and $\beta$, $80\sim 93\%$
 for $\Omega_m$). From Figs.~\ref{fig5} and~\ref{fig6}, it is easy to see
 that the uncertainties $\langle\sigma\rangle$ for $\alpha$, $\beta$ and
 $\Omega_m$ in the cases of $z_d=0.1$ are all much larger than (about 2
 times of) the ones in the cases of $z_d=0.2$ and 0.5. Thus, $z_d=0.1$ is
 not~suitable.~Although the constraints on $\alpha$ and $\beta$ in the
 cases of $z_d=0.5$ are better than the ones in the cases of $z_d=0.2$, the
 constraints on the cosmological parameter $\Omega_m$ are at the same level
 in both cases of $z_d=0.5$ and 0.2. In this sense, we prefer $z_d=0.2$ over
 0.5 since a lower redshift divide is easier to achieve, as mentioned
 above.~The number of type~Ib FRBs ($N_{\rm FRB}=300$ or 500) makes
 almost no difference at this point. So, we consider that $z_d=0.2$ might be
 a suitable choice on balance.


\subsection{The precision of the distance modulus}\label{sec4e}

In the fiducial case, we have assumed that the distance modulus $\mu$ can
 be measured with high precision, namely the relative error $\sigma_{\mu
 \cmn {\rm rel}}=0.2\%$~\cite{Spergel:2013} in the era of the Roman Space
 Telescope (formerly WFIRST). But before the launch of the Roman Space
 Telescope, how does this precision affect the calibration and
 the cosmological constraints?

We consider the relative error $\sigma_{\mu\cmn {\rm rel}}=0.2\%$, $0.5\%$
 and $1\%$ for the cases of $N_{\rm FRB}=300$ and 500, respectively, while
 $z_d=0.2$ and $\sigma_{F\cmn {\rm rel}}=\sigma_{S\cmn {\rm rel}}=1\%$ (as
 in the fiducial case). Again, in Figs.~\ref{fig7} and~\ref{fig8} we present
 the marginalized $1\sigma$ constraints on the parameters $\alpha$, $\beta$
 and $\Omega_m$ for 100 simulations in the cases of $N_{\rm FRB}=300$ and
 500, respectively.~We find from Figs.~\ref{fig7} and~\ref{fig8} that the
 constraints from the mock type~Ib FRBs are fairly reliable and robust,
 since the assumed values $\alpha=0.9$, $\beta=65$, and $\Omega_m=0.3153$
 used in Sec.~\ref{sec4a} to generate the mock type~Ib FRBs can be found
 within $1\sigma$ region in most of the 100 simulations (namely $62\sim
 77\%$ for $\alpha$ and $\beta$, $87\sim 97\%$ for $\Omega_m$).~From
 Figs.~\ref{fig7} and~\ref{fig8}, it is easy to see that the constraints on
 $\alpha$, $\beta$ and $\Omega_m$ become worse for larger $\sigma_{\mu
 \cmn {\rm rel}}$.~The number of type~Ib FRBs ($N_{\rm FRB}=300$ or 500)
 makes almost no difference at this point. In particular, the mean of
 the uncertainties $\langle\sigma\rangle=0.1724$ for $\Omega_m$ in the case
 of $N_{\rm FRB}=300$ and $\sigma_{\mu\cmn {\rm rel}}=1\%$ (bottom-right
 panel of Fig.~\ref{fig7}) is too large ($\sim 55\%$) compared with the
 assumed value $\Omega_m=0.3153$.~But it can decrease to $\langle\sigma
 \rangle=0.1016$ at the price of increasing the number of type~Ib FRBs
 to $N_{\rm FRB}=500$ (bottom-right panel of Fig.~\ref{fig8}). The case of
 $\sigma_{\mu\cmn {\rm rel}}=0.2\%$ is best, while the case of
 $\sigma_{\mu\cmn {\rm rel}}=0.5\%$ is also acceptable, regardless of
 $N_{\rm FRB}=300$ or 500. Note that currently SNIa can be measured with
 the precision around $0.6\%$ (see e.g.~Union2.1~\cite{Suzuki:2011},
 Pantheon~\cite{Pan-STARRS1:2017jku} and Pantheon+~\cite{Scolnic:2021amr,
 Brout:2022vxf} SNIa samples).~Thus, it is reasonable to expect
 $\sigma_{\mu\cmn {\rm rel}}\leq 0.5\%$ in the near future for type~Ib FRBs
 which share the same host galaxies with SNIa. On the~other hand, if the
 Roman Space Telescope (formerly WFIRST) could be launched in mid-2020s
 as~planed, $\sigma_{\mu\cmn {\rm rel}}=0.2\%$~\cite{Spergel:2013} will
 be easily achieved very soon.


\subsection{The precisions of the fluence and the flux}\label{sec4f}

In the fiducial case, we have assumed that the fluence $F_\nu$ and the flux
 $S_\nu$ could be measured with high precisions, namely the relative errors
 $\sigma_{F\cmn {\rm rel}}=\sigma_{S\cmn {\rm rel}}=1\%$. Of course, such
 high precisions cannot be achieved currently in the actual data of FRBs.
 Since FRBs have become a very promising and thriving field in astronomy and
 cosmology recently, many observational efforts have been dedicated to FRBs,
 and hence it is reasonable to expect some breakthroughs in the observations
 and the instruments in the future. Thus, it is of interest to see the
 effects of these precisions on the calibration of type~Ib FRBs and their
 cosmological constraints.


 \begin{center}
 \begin{figure}[tb]
 \centering
 \vspace{-9mm} \hspace{-6mm} 
 \includegraphics[width=0.82\textwidth]{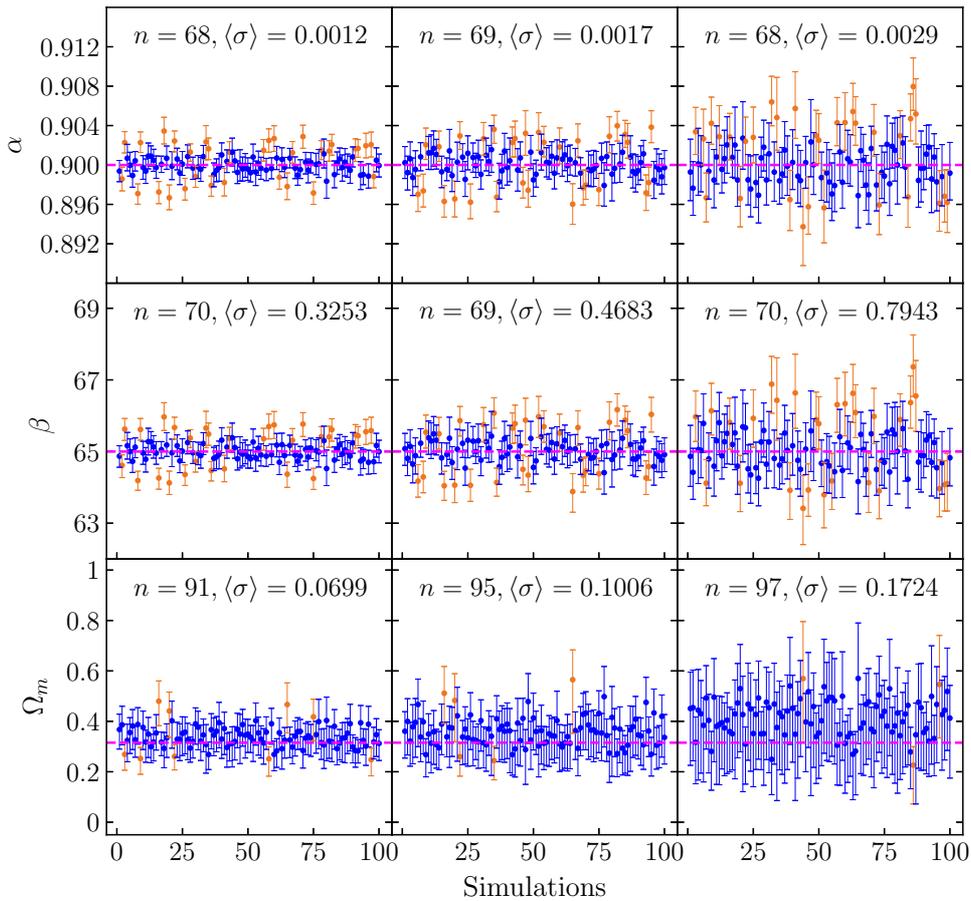}
 \vspace{-2mm} 
 \caption{\label{fig7} The same as in Fig.~\ref{fig3}, but for the cases
 of $N_{\rm FRB}=300$, $z_d=0.2$, and $\sigma_{\mu \cmn {\rm rel}}=0.2\%$
 (left panels), $0.5\%$ (middle panels), $1\%$ (right panels), respectively.
 See Sec.~\ref{sec4e} for details.}
 \end{figure}
 \end{center}



 \begin{center}
 \begin{figure}[tb]
 \centering
 \vspace{-9mm} \hspace{-6mm} 
 \includegraphics[width=0.82\textwidth]{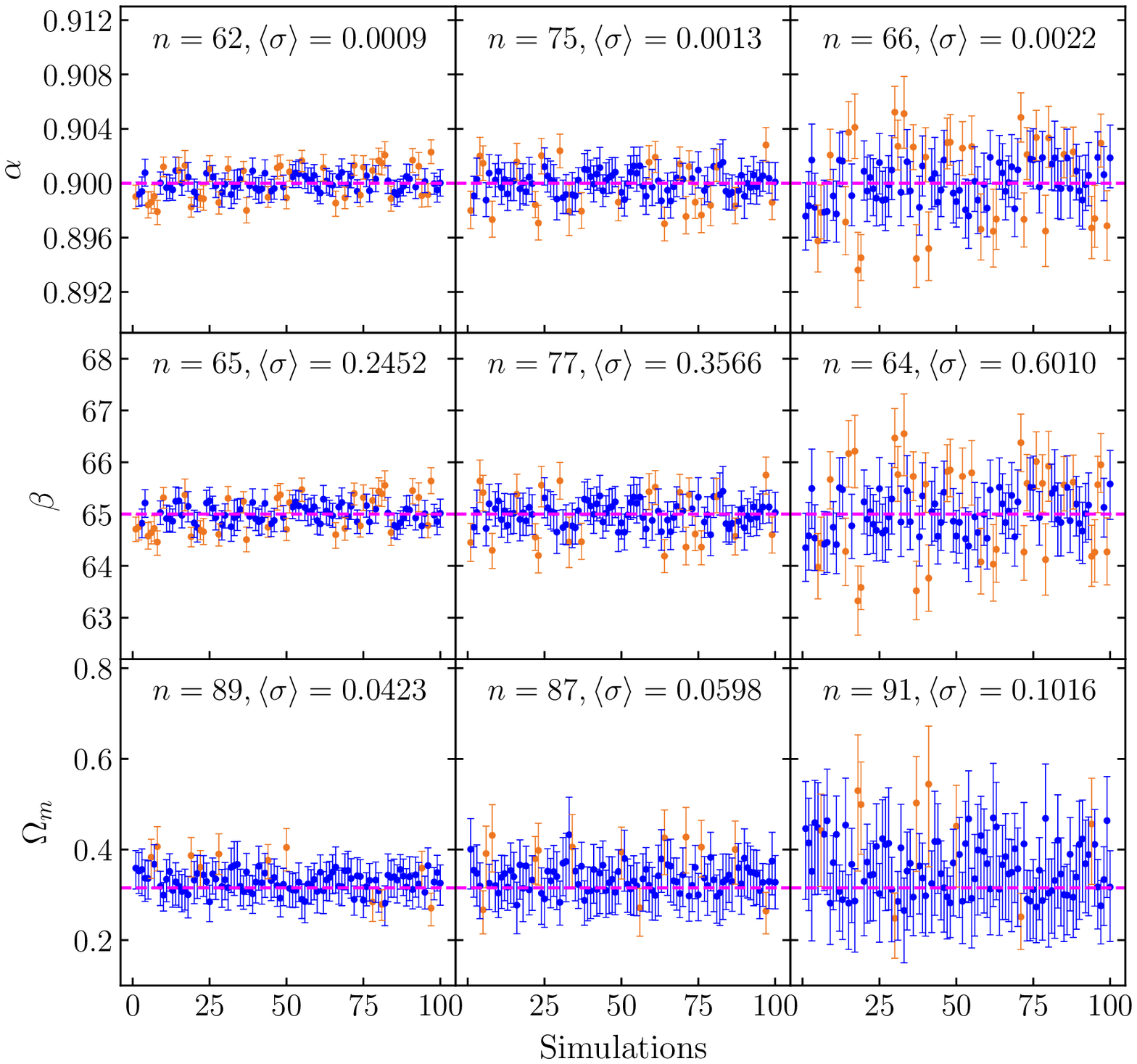}
 \vspace{-2mm} 
 \caption{\label{fig8} The same as in Fig.~\ref{fig3}, but for the cases
 of $N_{\rm FRB}=500$, $z_d=0.2$, and $\sigma_{\mu \cmn {\rm rel}}=0.2\%$
 (left panels), $0.5\%$ (middle panels), $1\%$ (right panels), respectively.
 See Sec.~\ref{sec4e} for details.}
 \end{figure}
 \end{center}



 \begin{center}
 \begin{figure}[tb]
 \centering
 \vspace{-8.3mm}\hspace{-6mm} 
 \includegraphics[width=0.82\textwidth]{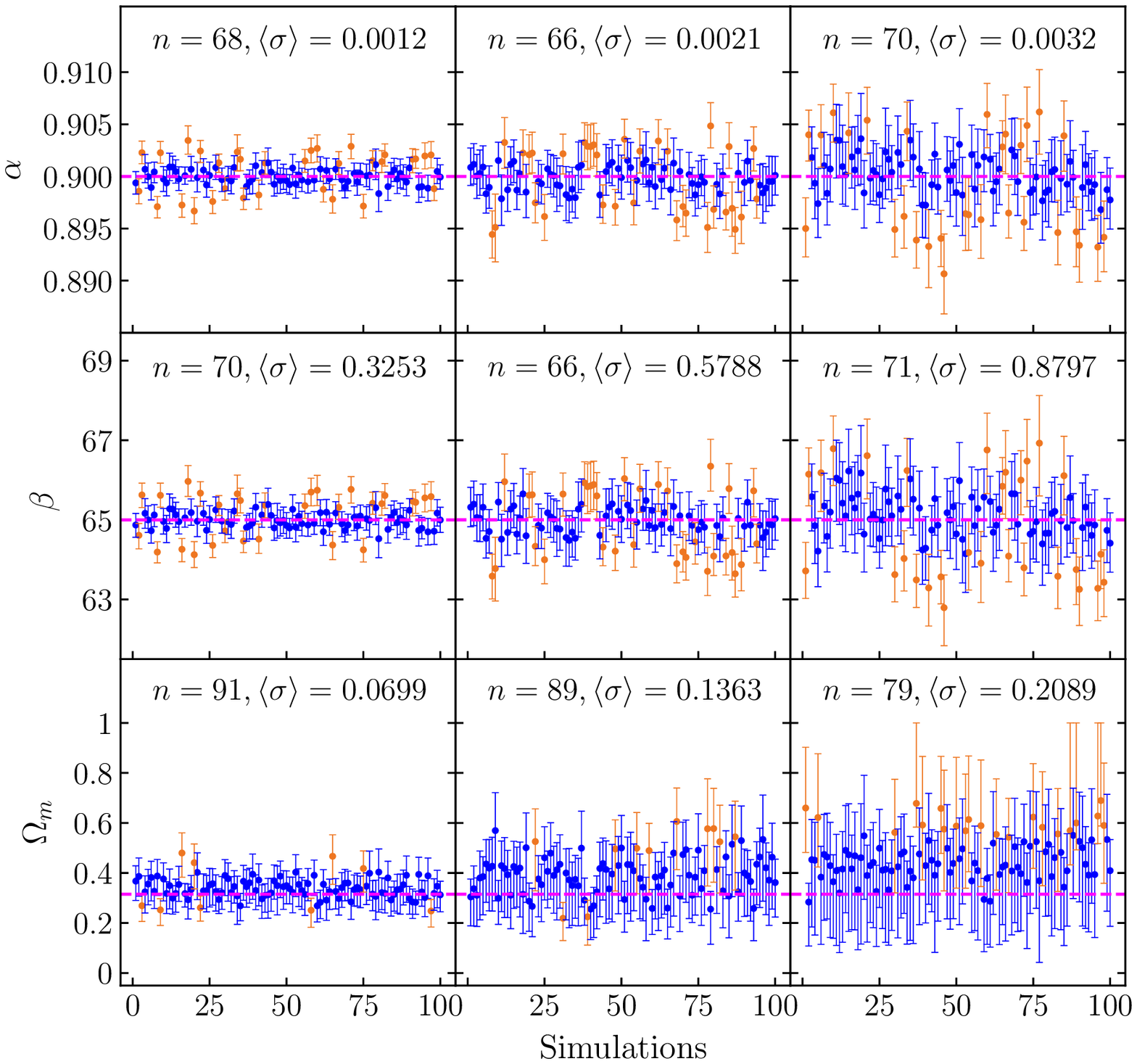}
 \vspace{-2mm} 
 \caption{\label{fig9} The same as in Fig.~\ref{fig3}, but for the cases
 of $N_{\rm FRB}=300$, $z_d=0.2$, $\sigma_{\mu \cmn {\rm rel}}=0.2\%$, and
 $\sigma_{F\cmn {\rm rel}}=\sigma_{S\cmn {\rm rel}}=1\%$ (left panels),
 $2\%$ (middle panels), $3\%$ (right panels), respectively. See
 Sec.~\ref{sec4f} for details.}
 \end{figure}
 \end{center}



 \begin{center}
 \begin{figure}[tb]
 \centering
 \vspace{-9.2mm}\hspace{-6mm} 
 \includegraphics[width=0.82\textwidth]{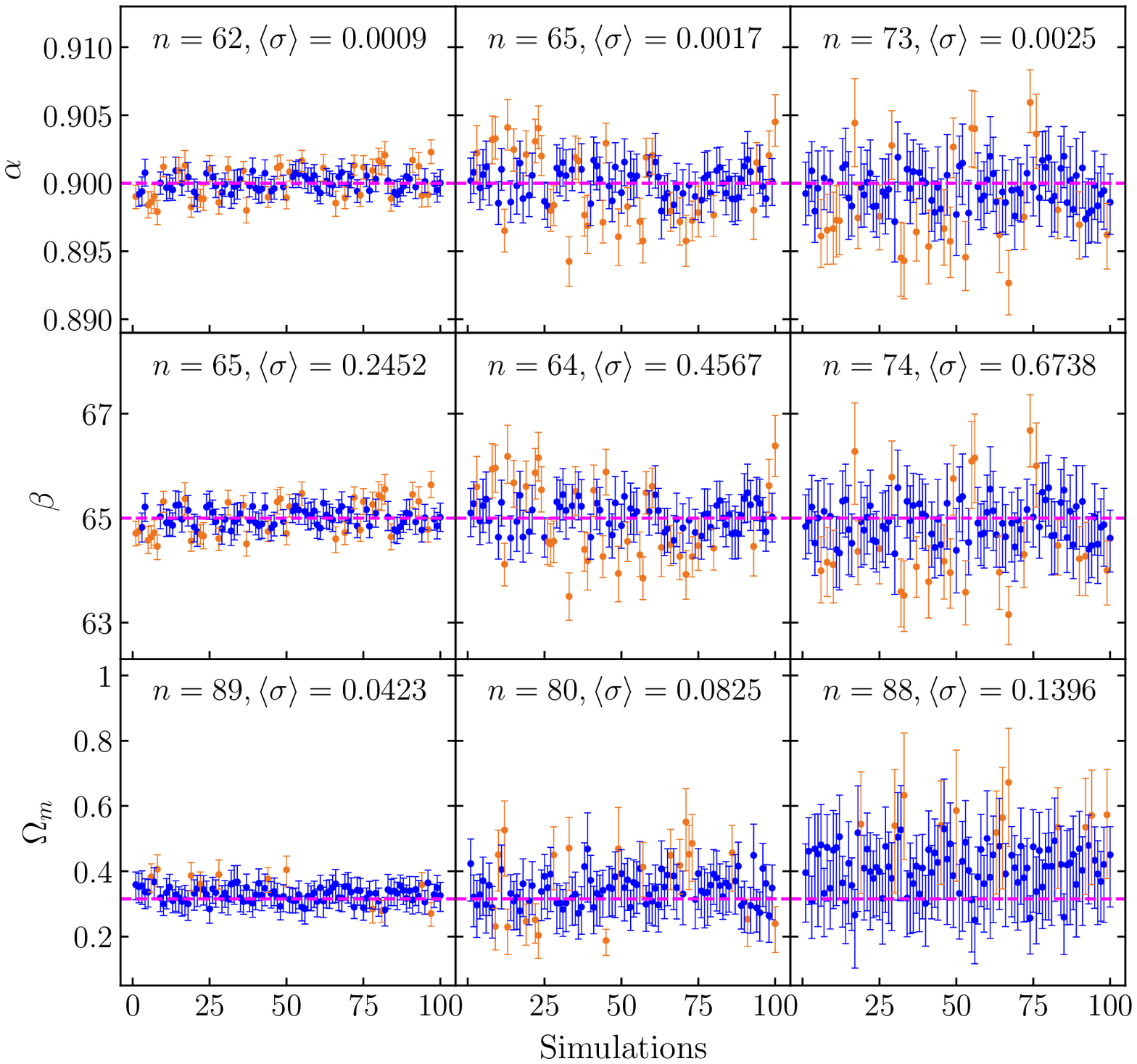}
 \vspace{-2mm} 
 \caption{\label{fig10} The same as in Fig.~\ref{fig3}, but for the cases
 of $N_{\rm FRB}=500$, $z_d=0.2$, $\sigma_{\mu \cmn {\rm rel}}=0.2\%$, and
 $\sigma_{F\cmn {\rm rel}}=\sigma_{S\cmn {\rm rel}}=1\%$ (left panels),
 $2\%$ (middle panels), $3\%$ (right panels), respectively. See
 Sec.~\ref{sec4f} for details.}
 \end{figure}
 \end{center}



 \begin{center}
 \begin{figure}[tb]
 \centering
 \vspace{-7mm} \hspace{-6mm} 
 \includegraphics[width=0.82\textwidth]{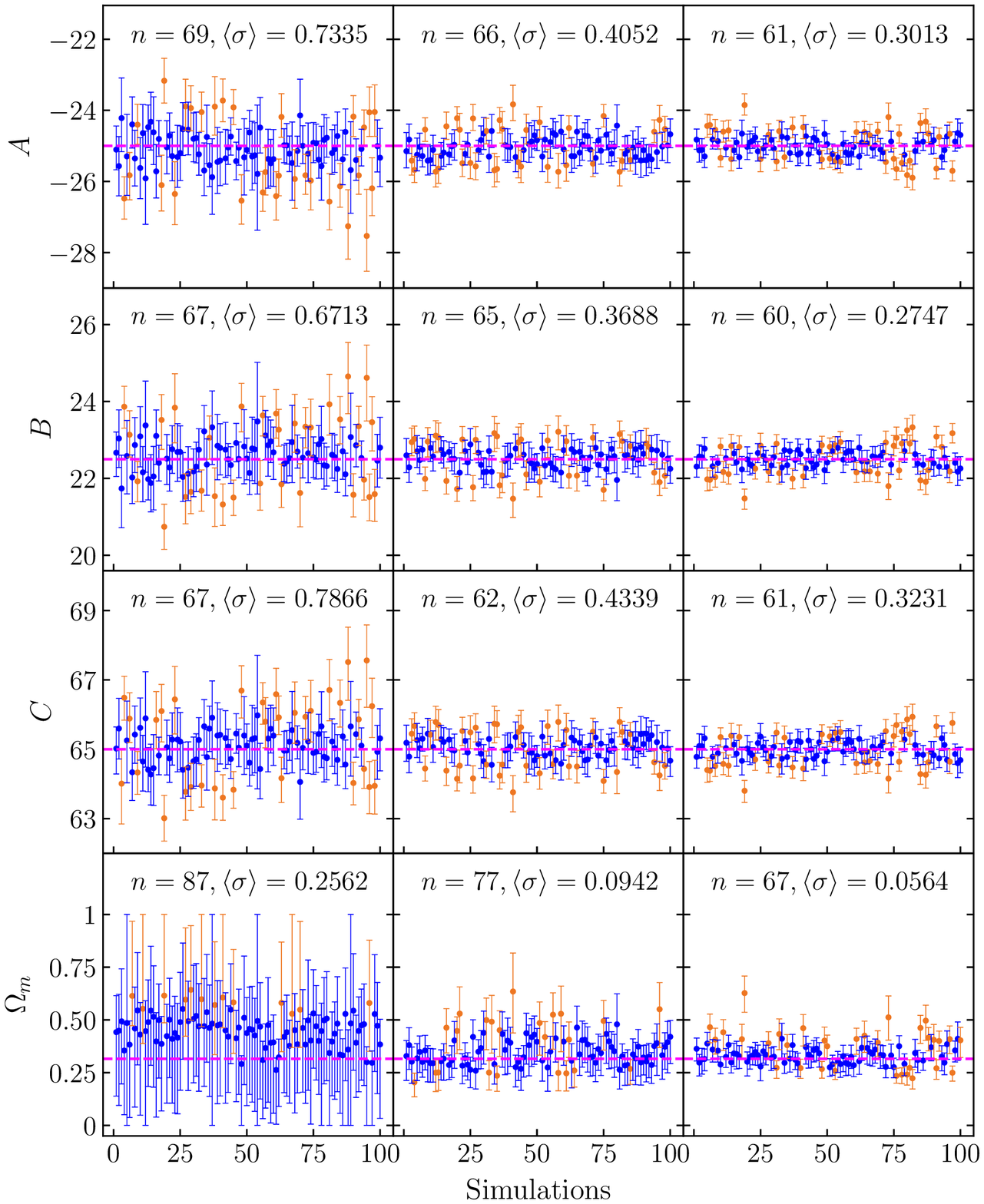}
 \vspace{-2mm} 
 \caption{\label{fig11} The same as in Fig.~\ref{fig3}, but for the case
 of three-parameter empirical relation with $N_{\rm FRB}=100$ (left panels),
 300 (middle panels), 500 (right panels), and $z_d=0.2$, $\sigma_{\mu
 \cmn {\rm rel}}=0.2\%$, $\sigma_{F\cmn {\rm rel}}=\sigma_{S\cmn {\rm
 rel}}=1\%$. See Sec.~\ref{sec4g} for details.}
 \end{figure}
 \end{center}


\vspace{-41.2mm} 

We consider the relative errors $\sigma_{F\cmn {\rm rel}}=\sigma_{S\cmn
 {\rm rel}}=1\%$, $2\%$ and $3\%$ for the cases of $N_{\rm FRB}=300$ and
 500, respectively, while $z_d=0.2$ and $\sigma_{\mu\cmn {\rm rel}}=0.2\%$
 (as in the fiducial case). Similarly, we present the results
 in Figs.~\ref{fig9} and~\ref{fig10}. We find from Figs.~\ref{fig9}
 and~\ref{fig10} that the constraints from the mock type~Ib FRBs are fairly
 reliable and robust, since the assumed values $\alpha=0.9$, $\beta=65$, and
 $\Omega_m=0.3153$ used in Sec.~\ref{sec4a} to generate the mock type~Ib
 FRBs can be found within $1\sigma$ region in most of the 100 simulations
 (namely $62\sim 74\%$ for $\alpha$ and $\beta$, $79\sim 91\%$
 for $\Omega_m$). From Figs.~\ref{fig9} and~\ref{fig10}, it is easy to see
 that the constraints on $\alpha$, $\beta$ and $\Omega_m$ become worse for
 larger $\sigma_{F\cmn {\rm rel}}=\sigma_{S\cmn {\rm rel}}$. The number of
 type~Ib FRBs ($N_{\rm FRB}=300$ or 500) makes almost no difference at this
 point. In particular, the mean of the uncertainties $\langle\sigma\rangle=
 0.2089$ for $\Omega_m$ in the case of $N_{\rm FRB}=300$ and $\sigma_{F\cmn
 {\rm rel}}=\sigma_{S\cmn {\rm rel}}=3\%$ (bottom-right panel
 of Fig.~\ref{fig9}) is too large ($\sim 66\%$) compared with the assumed
 value $\Omega_m=0.3153$. Although it can decrease to $\langle\sigma\rangle
 =0.1396$ at the price of increasing the number of type~Ib FRBs
 to $N_{\rm FRB}=500$ (bottom-right panel of Fig.~\ref{fig10}), this is
 still unacceptable since $0.1396/0.3153\simeq 44\%$. In this sense, the
 case of $\sigma_{F\cmn {\rm rel}}=\sigma_{S\cmn {\rm rel}}=3\%$ is not
 enough to get the acceptable constraints on the cosmological models. One of
 the ways out is to significantly increase the number of type~Ib FRBs (say,
 $N_{\rm FRB}=1000$ or even more). On the other hand, the cases
 of $\sigma_{F\cmn {\rm rel}}=\sigma_{S\cmn {\rm rel}}=1\%$ and $2\%$ are
 acceptable. Anyway, improving the precisions of the fluences and the fluxes
 for FRBs is an important task in the future.


\subsection{Three-parameter empirical relation}\label{sec4g}

In the above discussions, the empirical relation given in
 Eq.~(\ref{eq14}) is considered. This is a two-parameter empirical relation,
 with $\alpha$ and $\beta$ as the free parameters.~The coefficients for the
 terms of $\hspace{0.05em}\log F_\nu$ and $\log S_\nu$ are not independent.
 The cause roots in the empirical $L_\nu-E$ relation given in
 Eq.~(\ref{eq13}) due to~the unknown physical mechanism for the origins
 of type~Ib FRBs.

However, if one pretends to do not know the empirical $L_\nu-E$ relation
 given in Eq.~(\ref{eq13}), the empirical relation between the distance
 modulus $\mu$ (equivalently the luminosity distance $d_L$), the fluence
 $F_\nu$ and the flux $S_\nu$ could be instead regarded as a three-parameter
 empirical relation without any underlying physical mechanism for the
 origins of type~Ib FRBs, namely
 \be{eq23}
 \mu=A\log\frac{\;F_\nu/(1+z)\,}{\rm Jy\;ms}
 +B\log\frac{S_\nu}{\;\rm Jy\,}+C\,,
 \ee
 where the independent parameters $A$, $B$ and $C$ are all dimensionless
 constants. The assumed underlying empirical relation given by
 Eq.~(\ref{eq20}) used in Sec.~\ref{sec4a} to generate the mock type~Ib
 FRBs corresponds to $A=-25$, $B=22.5$ and $C=65$. One can calibrate the
 three-parameter empirical relation~(\ref{eq23}) by using type~Ib FRBs
 at low redshifts $z<z_d$, and then obtain the distance moduli $\mu$ for
 type~Ib FRBs at high redshifts $z\geq z_d$ by using this calibrated
 three-parameter empirical relation. Thus, one can constrain
 the cosmological models with these type~Ib FRBs at high
 redshifts $z\geq z_d$.

We consider the cases of $N_{\rm FRB}=100$, 300 and 500 with $z_d=0.2$,
 $\sigma_{\mu\cmn {\rm rel}}=0.2\%$, and $\sigma_{F\cmn {\rm rel}}=
 \sigma_{S\cmn {\rm rel}}=1\%$ (as in the fiducial case).~In
 Fig.~\ref{fig11}, we present the marginalized $1\sigma$ constraints on
 the parameters $A$, $B$, $C$ and $\Omega_m$ for 100 simulations in the
 cases of $N_{\rm FRB}=100$, 300 and 500, respectively.~We find from
 Fig.~\ref{fig11} that the constraints from the mock type~Ib FRBs are
 fairly reliable and robust, since the assumed values $A=-25$, $B=22.5$,
 $C=65$, and $\Omega_m=0.3153$ used in Sec.~\ref{sec4a} to generate the
 mock type~Ib FRBs can be found within $1\sigma$ region in most of the
 100 simulations (namely $60\sim 69\%$ for $A$, $B$ and $C$, $67\sim
 87\%$ for $\Omega_m$).~Comparing Fig.~\ref{fig11} with the cases of
 two-parameter empirical relation, the uncertainties $\langle\sigma\rangle$
 become larger. This is not surprising, since the number of independent
 parameters has been increased.~On the other hand, the mean of
 the uncertainties $\langle\sigma\rangle=0.2562$ for $\Omega_m$ in the case
 of $N_{\rm FRB}=100$ (bottom-left panel of Fig.~\ref{fig11}) is too
 large ($\sim 81\%$) compared with the assumed value $\Omega_m=
 0.3153$.~Thus, $N_{\rm FRB}=100$ is not enough to get the acceptable
 constraints.~But the uncertainties dramatically decrease in the cases
 of $N_{\rm FRB}=300$ and 500, and hence they are acceptable.~Comparing
 with the middle panels of Fig.~\ref{fig5} ($N_{\rm FRB}=300$)
 and Fig.~\ref{fig6} ($N_{\rm FRB}=500$) in the case of two-parameter
 empirical relation, we find that the constraints on the cosmological
 parameter $\Omega_m$ become looser in the case of three-parameter empirical
 relation. So, we consider that the two-parameter
 empirical relation~(\ref{eq14}) should be preferred in the FRB cosmology.


\section{Concluding remarks}\label{sec5}

Recently, FRBs have become a thriving field in astronomy and cosmology. Due
 to their extragalactic and cosmological origin, they are useful to study
 the cosmic expansion and IGM. In the literature, the dispersion measure DM
 of FRB has been considered extensively. It could be used as an indirect
 proxy of the luminosity distance $d_L$ of FRB. The observed DM contains the
 contributions from the Milky Way (MW), the MW halo, IGM, and the host
 galaxy.~Unfortunately, IGM and the host galaxy of FRB are poorly known
 to date, and hence the large uncertainties of $\rm DM_{IGM}$
 and $\rm DM_{host}$ in DM plague the FRB cosmology. Could we avoid
 DM in studying cosmology? Could we instead consider the luminosity distance
 $d_L$ directly in the FRB cosmology? We are interested to find a way
 out for this problem in the present work. From the lessons of
 calibrating SNIa or long GRBs as standard candles, we consider a universal
 subclassification scheme for FRBs, and there are some empirical relations
 for them. In the present work, we propose to calibrate type~Ib FRBs
 as standard candles by using a tight empirical relation without DM.
 The calibrated type~Ib FRBs at high redshifts can be used like SNIa
 to constrain the cosmological models. We also test the key factors
 affecting the calibration and the cosmological constraints.

We find that the constraints become better for the larger number of type~Ib
 FRBs. $N_{\rm FRB}=100$ is not enough to get the acceptable constraints,
 while $N_{\rm FRB}=300$ or 500 are suitable.~We also find that the redshift
 divide $z_d=0.1$ is not suitable, and we suggest that $z_d=0.2$ might be a
 suitable choice on balance.~It is found that the precision of the distance
 modulus $\sigma_{\mu\cmn {\rm rel}}\leq 0.5\%$ is enough to get the
 acceptable constraints, which can be achieved very soon.~On the other hand,
 the precisions of the fluence and the flux $\sigma_{F\cmn {\rm rel}}=
 \sigma_{S\cmn {\rm rel}}\geq 3\%$ is not enough to get the acceptable
 constraints on the cosmological models.~One of the ways out is
 to significantly increase the number of type~Ib FRBs (say, $N_{\rm FRB}=
 1000$ or even more).~Anyway, improving the precisions of the fluences and
 the fluxes for FRBs is an important task in the future.~Although one could
 instead consider a three-parameter empirical relation~(\ref{eq23}), we
 suggest that the two-parameter empirical relation~(\ref{eq14}) should be
 preferred in the FRB cosmology.

Of course, current data of FRBs are certainly not enough to calibrate
 type~Ib FRBs as standard candles. So, it is a proof of concept by using the
 simulated FRBs in the present work. But the same pipeline holds for the
 actual type~Ib FRBs in the future. Since FRBs have become a very promising
 and thriving field in astronomy and cosmology recently, many observational
 efforts have been dedicated to FRBs, and hence it is reasonable to expect
 some breakthroughs in the observations and the instruments in the future.
 Let us be optimistic with the hope to actually use type~Ib FRBs as standard
 candles.

Clearly, the key to calibrate type~Ib FRBs as standard candles is that there
 is really a tight empirical relation for them, due to the unknown physical
 mechanism for the origins of type~Ib FRBs. The universal subclassification
 scheme for FRBs proposed in~\cite{Guo:2022wpf} and the empirical relations
 found in~\cite{Guo:2022wpf} should be carefully examined by using the
 larger and better FRB datasets in the future.~On the other hand, the
 theories that could produce such an empirical relation for type~Ib FRBs
 are desirable.

Notice that in the present work we constrain the cosmological model by
 using the calibrated type~Ib FRBs at high redshifts alone. But this is not
 necessary. In fact, similar to the case of SNIa, it is better to combine
 them with other observations (e.g.~cosmic microwave background~(CMB)
 and large-scale structure~(LSS)) to obtain
 the much tighter cosmological constraints.

In the present work, the subclasses of FRBs are called type I, II, a, b,
 Ia, Ib, IIa, IIb FRBs as in~\cite{Guo:2022wpf}, respectively.
 However, these terms might be hard to remember. So, the alternative
 terms nFRBs, rFRBs, oFRBs, yFRBs, noFRBs, nyFRBs, roFRBs, ryFRBs are
 suggested in~\cite{Li:2024dge}, which might be friendly and easy to
 remember. We refer to~\cite{Li:2024dge} for more details.

Several years passed after the first CHIME/FRB
 catalog~\cite{CHIMEFRB:2021srp}. To date, more than 50 FRBs
 have been well localized, and hence their redshifts $z$ are
 observationally known. In~\cite{Li:2024dge}, the empirical relations
 have been carefully checked with the actual data of current
 localized FRBs.~It has been found in~\cite{Li:2024dge} that
 many empirical relations for FRBs still hold.~In particular,
 the empirical $L_\nu-E$ relation used here to calibrate FRBs
 as standard candles for cosmology stands firm with the current
 $44\sim 52$ localized FRBs, and its slope $a$ and intercept
 $b$ obtained from the actual data are fairly close to the ones
 in Eq.~(\ref{eq9}). Note that the uncertainties are also taken into
 account in~\cite{Li:2024dge}. It is found that the slope $a\not=1$ in
 the empirical $L_\nu-E$ relation~(\ref{eq13}) far beyond $3\sigma$
 confidence level (C.L.), actually on the edge of $4\sigma$
 C.L.~\cite{Li:2024dge}. If $a=1$, the luminosity distance $d_L$ will be
 canceled in both sides of Eq.~(\ref{eq13}) (n.b.~Eq.~(\ref{eq8})), and
 then it cannot be used to study cosmology. But actually $a\not=1$ as
 shown in~\cite{Li:2024dge} by using the current localized FRBs, and
 hence it supports the empirical $L_\nu-E$ relation~(\ref{eq13}) used
 here to calibrate FRBs as standard candles for cosmology. We strongly
 refer to~\cite{Li:2024dge} for more details.


\section*{ACKNOWLEDGEMENTS}

We thank the anonymous referee for quite useful comments
 and suggestions, which helped us to improve

\newpage 

\noindent this work.~We are grateful
 to Da-Chun~Qiang, Hua-Kai~Deng, Shupeng~Song, Jing-Yi~Jia, Shu-Ling~Li,
 Yun-Long~Wang, Lin-Yu~Li and Jia-Lei~Niu for kind help and
 useful discussions. This work was supported in part by
 NSFC under Grants No.~12375042, No.~11975046 and No.~11575022.

\renewcommand{\baselinestretch}{1.127}


\end{document}